\documentclass [prb,reprint,notitlepage,superscriptaddress,floatfix] {revtex4-2}
\usepackage{amsmath}
\usepackage{graphicx}
\usepackage{lmodern}
\usepackage{amsmath}
\usepackage{simpler-wick}
\usepackage[caption=false]{subfig}
\usepackage[colorlinks=true, citecolor=green, urlcolor=blue, linkcolor=red]{hyperref}
\usepackage{color}
\usepackage{physics}

\usepackage{amssymb}
\newcommand{\be}{\begin{equation}}
\newcommand{\ee}{\end{equation}}
\renewcommand{\(}{\left(}
\renewcommand{\)}{\right)}
\renewcommand{\[}{\left[}
\renewcommand{\]}{\right]}

\DeclareMathOperator{\sgn}{sgn}

\newcommand{\da}{^\dagger}
\newcommand{\pd}{\partial}
\newcommand{\ep}{\varepsilon}

\begin{document}

\title {Quantum chaos and phase transition in the Yukawa-SYK model}
\author{Andrew Davis}
\author{Yuxuan Wang}
\affiliation{Department of Physics, University of Florida, Gainesville, Florida, 32601, USA}
\date{\today}
\begin{abstract}
We analyze the relation between quantum chaotic behavior and phase transition of the Yukawa-SYK model as a function of filling and temperature, which describes random Yukawa interactions between $N$ complex fermions and $M$ bosons in zero spatial dimensions, for both the non-Fermi liquid and insulating states at finite temperature and chemical potential. We solve the ladder equations for the out-of-time-order correlator (OTOC) for both the bosons and fermions. Despite the appearance of the chemical potential in the Hamiltonian, which explicitly introduces an additional energy scale, the OTOCs for the fermions and bosons in the non-Fermi liquid state turn out to be unaffected, and the Lyapunov exponents that diagnose chaos remain maximal. As the chemical potential increases, the system is known to experience a first-order transition from a critical phase to a gapped insulating phase. We postulate that the boundary of the region in parameter space where each phase is (meta)stable coincides with the curve on which the Lyapunov exponent is maximal. By calculating the exponent in the insulating phase and comparing to numerical results on the boundaries of stability, we show that this is plausible.
\end{abstract}
\maketitle

\section{Introduction}

The Sachdev-Ye-Kitaev (SYK) model~\cite{kitaevsyk} has attracted great interest in both high-energy and condensed matter communities~\cite{Chowdhury:2021qpy,PhysRevD.95.026009,Rosenhaus:2018dtp}. It consists of $N$ flavors of fermions in $(0+1)$-d coupled by a random, Gaussian, all-to-all interaction, and it exhibits several remarkable properties~\cite{PhysRevD.94.106002}. First, it is exactly solvable in the large-$N$ limit, despite the strong coupling. Second, it possesses a (near) conformal symmetry in the infrared and has a holographic dual, corresponding to a black hole in near-AdS$_2$, making it a tractable platform for the study of AdS/CFT correspondence~\cite{Kitaev:2017awl}. Finally, it is a fast scrambler -- it is maximally chaotic as measured by the Lyapunov exponent $\lambda_L=2\pi T$, a feature shared by black holes in Einstein gravity~\cite{PhysRevD.94.106002}. This makes it noteworthy as an example of a solvable system for quantum gravity.

On the condensed matter side, the SYK model and its variants are useful platforms for the understanding of non-Fermi liquid (nFL) physics~\cite{2022arXiv220304990P,PhysRevB.103.235129,PhysRevB.105.235111}, which is relevant to the strange metal behavior of unconventional superconductors~\cite{PhysRevB.57.5505,RevModPhys.73.797,2014arXiv1409.4673K,PhysRevLett.63.1996,Cha18341,Guo:2020aog}. Non-Fermi liquids are characterized by a power-law frequency dependence of the fermion self-energy and a resulting spectral function without quasiparticle peaks, and occurs, for example, when massless bosonic modes destroy the coherence of the fermions~\cite{doi:10.1146/annurev-conmatphys-031016-025531}.
While the fermions in the SYK model display nFL behavior in the near-conformal regime, perhaps more relevant to condensed matter physics is a variant of the SYK model, dubbed the Yukawa-SYK (Yukawa-SYK) model, in which the random four-fermion interaction is replaced with a random Yukawa interaction between $N$ flavors of complex fermions and $M$ flavors of bosons with bare mass $m_0$. The Yukawa-SYK model has been studied in the context of superconducting instabilities of nFLs~\cite{Wang2020SolvableSQ, PhysRevB.100.115132,2022ScPP...12..151C,HAUCK2020168120} and has been generalized to finite dimensions~\cite{PhysRevB.103.L081113}. Interestingly, it has been shown that the pairing transition is dual to that of a holographic superconductor in AdS$_2$ spacetime~\cite{SchmalianHolographic2022}.

At a generic filling factor, the Yukawa-SYK model has a nFL phase as well as an incompressible, insulating phase.~\cite{Wang:2020dtj}
The nFL state exhibits power-law self-energies for both the fermions and bosons with $M/N$-dependent exponents, and also has the interesting property that the boson mass flows to zero even if the bare mass is large (this ``self-tuning'' also occurs in $(1+1)$-d but not higher dimensions~\cite{PhysRevB.103.L081113}). For small chemical potential, the nFL is stable, but at larger $\mu$ there are nFL solutions with negative compressibility, indicating an unstable phase. At a critical value of $\mu$, there is a first-order transition to the insulating phase.
The phase diagram of the model has been studied analytically and numerically in~\cite{PhysRevResearch.3.013250, PhysRevB.103.195108}. It is worth noting that the original SYK model with complex fermions also displays a similar phase diagram~\cite{PhysRevLett.120.061602,PhysRevResearch.3.033089}.


At half filling with $\mu=0$, the Yukawa-SYK model is known to be maximally chaotic~\cite{PhysRevB.101.125112}. In this work, we focus on the quantum chaotic properties of the Yukawa-SYK model in $(0+1)$-d away from half filling as a function of chemical potential $\mu$ and their relation to the phase transition. It is typical to quantify the rate at which chaos develops by the Lyapunov exponent $\lambda_L$~\cite{Stanford:2015owe, PhysRevB.101.125112, PhysRevB.103.L081113,Gu:2016oyy}. In classical chaotic systems, even slight differences in initial conditions result in trajectories $q(t)$ that diverge exponentially: $\delta q(t)/\delta q(0) \propto e^{\lambda_L t}$. In quantum many-body systems, the analogous object is an out-of-time-order correlator (OTOC)~\cite{Stanford:2015owe} which also grows exponentially and defines a quantum Lyapunov exponent (correlators of this form had been considered earlier by~\cite{1969JETP...28.1200L}). We will compute the 
following correlator for the fermions:
\begin{align}
F_{c}(t_1,t_2) \sim
\Tr[\rho^\frac{1}{2} \{c(t_1), c\da(0)\} \rho^\frac{1}{2} \{c(t_2), c\da(0)\}\da],
\label{eq:1}
\end{align}
When $t_1=t_2=t$, this defines the OTOC, which involves the trace of a positive semidefinite operator and ensures a real Lyapunov spectrum. The Lyapunov exponent can then be extracted from the leading exponential growth of this OTOC, $F_{c}(t_1,t_2)\sim \exp[\lambda_L(t_1+t_2)/2]$. It has been argued~\cite{Maldacena:2015waa} that in quantum systems at temperature $T$ there is an upper bound~\cite{Maldacena:2015waa} on the Lyapunov exponent $\lambda_L \leq 2\pi k_B T/\hbar$ (we set $k_B=\hbar=1$ henceforth), which is known to be saturated by black holes and SYK-like models~\cite{PhysRevD.94.106002}.

The key result of this work is that $\lambda_L=2\pi T$ to leading order 
for the entire nFL phase; on the other hand, $\lambda_L$ is exponentially suppressed in the insulating phase. As was mentioned, the two phases are separated by a first-order phase transition, which means there is a hysteresis regime in $(\mu, T)$-space where both the nFL and insulating solutions are stable or metastable~\cite{PhysRevB.103.195108}, and the transition occurs somewhere in this region when the global minimum of the free energy switches from one solution to the other. From the upper bound for $\lambda_L$, we conjecture that the boundaries of this hysteresis region are curves on which the $\lambda_L$ for one of the phases attains its maximum value. Indeed, we numerically verify that the curve on which the insulating solution disappears matches very well with the curve on which $\lambda_L$ approaches $2\pi T$ for the insulating solution at low temperatures. For the nFL solution, since $\lambda_L$ universally approaches the upper bound to leading order, one needs to go beyond the conformal limit to compute the correction to $\lambda_L$ and to verify the conjecture. We leave this to a future study.

The rest of this paper is organized as follows. In Section~\ref{sec:2} we present the Yukawa-SYK model and review the known results at zero temperature for both the nFL and insulating states. We then map the zero temperature nFL Green's functions to finite temperature using the conformal symmetry of the model, and we calculate the various Green's functions that will be required for the perturbative expansion of the OTOC. In Section~\ref{sec:3} we show how the OTOC can be computed by summing a series of ladder diagrams and solve the resulting eigenfunction problem in detail. 
Finally, in Section~\ref{sec:4} we discuss the implications of our results for the OTOC on the phase diagram of the Yukawa-SYK model and compare with some known numerical results.

\section{Model and preliminaries}
\label{sec:2}

The Yukawa-SYK model~\cite{Wang2020SolvableSQ, PhysRevB.100.115132,2022ScPP...12..151C,HAUCK2020168120,Wang:2020dtj} consists of $M$ flavors of 
complex fermions and $N$ flavors of bosons interacting via a Yukawa term with a random coupling. The bosons have bare mass $m_0$; the fermions are dispersionless, and the fermion density is controlled by the chemical potential $\mu$; see Refs.~\cite{quiver-1,quiver-2} for related works on a similar model in string-inspired quiver quantum mechanics. The Lagrangian of the system in imaginary time is
\begin{align}
L =& \sum_i^M \qty[c_i\da (\pd_\tau - \mu) c_i]
+ \sum_\alpha^N \qty[\frac{1}{2}(\pd_\tau \phi_\alpha)^2 + \frac{1}{2}m_0^2 \phi_\alpha^2] \nonumber\\
&+\frac{i}{\sqrt{MN}}\sum_{ij\alpha} t^\alpha_{ij} c_i\da c_j \phi_\alpha.
\end{align}
For each realization of the system, the coupling is drawn from a Gaussian distribution with $\expval{t^\alpha_{ij}}=0$ and $\expval{t^\alpha_{ij}t^\beta_{kl}}=\omega_0^3(\delta_{ik}\delta_{jl} - \delta_{il}\delta_{jk})\delta_{\alpha\beta}$, with $\omega_0 > 0$. For the rest of this work, we assume the ``weak-coupling limit" $\omega_0\ll m_0$, in which, despite the name, the theory remains non-perturbative. Without loss of generality we can restrict our attention to $\mu > 0$ because of particle-hole symmetry.

This model has already been analyzed at zero temperature in~\cite{Wang2020SolvableSQ, PhysRevB.100.115132,2022ScPP...12..151C,HAUCK2020168120,Wang:2020dtj} and at finite temperature in~\cite{PhysRevB.103.195108}. We assume a replica-diagonal solution and define the fermion and boson self-energies through $G^{-1}(\omega) = i\omega + \mu + \Sigma(\omega)$ and $D^{-1}(\Omega)= \Omega^2 + m_0^2 + \Pi(\Omega)$. The Schwinger-Dyson (SD) equations in the imaginary time domain are then
\begin{align}
\label{SDa}
&\Sigma(\tau, \tau') = -\omega_0^3 D(\tau, \tau') G(\tau, \tau') \\
\label{SDb}
&\Pi(\tau, \tau') = \frac{M}{N} \omega_0^3 G(\tau, \tau') G(\tau', \tau)
\end{align}
It was shown that in the weak-coupling limit, 
the only important energy scales are $\mu$ and the ratio 
\begin{align}
\omega_F \equiv \omega_0^3/m_0^2.
\end{align}
 For $\mu > \omega_F/2$, the SD equation has a (meta)stable solution corresponding to an incompressible state: the filling $\nu = 1$ regardless of $\mu$, the bosonic self-energy vanishes, and the fermionic self-energy is a constant. On the other hand, for small $\mu$, the SD equations admit a non-Fermi liquid solution in which the Green's functions have power-law forms. We will compute the OTOC in both of these phases, beginning by deriving the various finite-temperature Green's functions that will appear in the expansion of the OTOC.

\subsection{Non-Fermi liquid state}
At zero temperature, for sufficiently small chemical potential with $\mu\lesssim \omega_F$, a solution of the Schwinger-Dyson equation, which is stable for $\mu\ll \omega_F$ and metastable for $\mu\sim \omega_F$, is\begin{align}
\label{conformalselfenergies}
\Sigma(\omega) = -\mu + \omega_f^{1-x}|\omega|^x (\alpha+i \sgn \omega) \equiv -\mu+\tilde\Sigma(\omega)\\
\Pi(\Omega) = -m_0^2 + \beta m_0^2 \omega_f^{2x-1} |\Omega|^{1-2x} \equiv -m_0^2 + \tilde\Pi(\Omega)
\end{align} 
where $\omega_f$ (related to but not to be confused with $\omega_F$) is a dynamically generated energy scale below which these solutions are valid. The parameters $\omega_f$, $\alpha$, $\beta$, and $x$ are related to the values of $\omega_0$, $m_0$, $\omega_F$, $\mu$, and $M/N$. The expressions for these parameters have been obtained in~\cite{Wang:2020dtj}, but, as we shall see, they will not enter the Lyapunov exponent. In particular we will make use of the relations
\begin{align}
\label{parameterIdentities}
\frac{M}{N} = \frac{1-2x}{x}
\frac{\cos(\pi x)}{\cos(\pi x)+\frac{1-\alpha^2}{1+\alpha^2}}, \nonumber\\
\frac{\omega_0^3}{4\pi\beta m_0^2\omega_f(1+\alpha^2)} = \frac{-\Gamma(-2x)}{\Gamma^2(-x)}.
\end{align}
In fact, the self-energies always have this form at sufficiently low frequencies regardless of $\omega_F$ and $m_0$ (even in the strong coupling limit $\omega_0\gg m_0$). 
In the conformal limit in which $i\omega \ll \Sigma(\omega)$ and $\Omega^2 \ll \Pi(\Omega)$, 
the Green's functions have the power-law forms $G(\omega) = \tilde\Sigma(\omega)^{-1}$ and $D(\Omega) = \tilde\Pi(\Omega)^{-1}$, which can be recast in the time domain as
\begin{align}
\label{GSigma}
\int \dd{\tau'} G(\tau, \tau') {\Sigma}(\tau', \tau'') =& \; \delta(\tau - \tau'') \\
\label{DPi}
\int \dd{\tau'} D(\tau, \tau') {\Pi}(\tau', \tau'') =& \; \delta(\tau - \tau'').
\end{align}
We have further dropped the tildes on the self-energies. In doing so, we are neglecting a contribution on the LHS of~\eqref{GSigma} which looks like $\int \dd{\tau'} G(\tau, \tau') \mu \delta(\tau' - \tau'')$ (and an \mbox{analogous} contribution in~\eqref{DPi}). This is permissible because it is only nonzero at infinitely short time scales, whereas we are interested in the low-energy (long-time) physics.

In the frequency domain, the Green's functions are
\begin{align}
G(\omega) =& \frac{\omega_f^{x-1}}{\alpha^2+1} \frac{\alpha -i \sgn \omega}{|\omega|^x} \\
D(\Omega) =& \beta^{-1} m_0^{-2} \omega_f^{1-2x} \frac{1}{|\Omega|^{1-2x}}.
\end{align}
Using the Fourier transform identities
\begin{align}
\int \frac{\dd{\omega}}{2\pi} e^{i \omega \tau} \frac{1}{|\omega|^{2\Delta}}
= &\frac{\sin(\pi\Delta)}{\pi} \Gamma(1-2\Delta) |\tau|^{2\Delta-1} \\
\int \frac{\dd{\omega}}{2\pi} e^{i \omega \tau} \frac{\sgn\omega}{|\omega|^{2\Delta}}
= & i \frac{\cos(\pi\Delta)}{\pi} \Gamma(1-2\Delta) |\tau|^{2\Delta-1} \sgn\tau,
\end{align}
we find in the time domain
\begin{align}
&G(\tau-\tau') =- A_G \sgn(\tau-\tau') \nonumber\\
&\times\cos(\frac{\pi x}{2} +\sgn(\tau-\tau')\arctan\alpha) \frac{1}{|\tau-\tau'|^{1-x}} \\
&D(\tau-\tau') =  A_D \frac{1}{|\tau-\tau'|^{2x}}
\end{align}
where $A_G = \omega_f^{x-1}(\alpha^2+1)^{-1/2}\pi^{-1}\Gamma(1-x)$ and $A_D \equiv \beta^{-1} m_0^{-2} \omega_f^{1-2x} \pi^{-1} \cos(\pi x) \Gamma(2x)$ are constants.

Taken together, equations~\eqref{SDa},~\eqref{SDb},~\eqref{GSigma}, and~\eqref{DPi} are invariant under the transformation $\tau\to f(\tau)$ and
\begin{align}
\label{Gmapping}
&G(\tau, \tau') \rightarrow [f'(\tau) f'(\tau')]^\Delta \frac{g(\tau)}{g(\tau')} G(f(\tau), f(\tau')) \\
&\Sigma(\tau, \tau') \rightarrow [f'(\tau) f'(\tau')]^{1-\Delta} \frac{g(\tau)}{g(\tau')} \Sigma(f(\tau), f(\tau')) \\
\label{Dmapping}
&D(\tau, \tau') \rightarrow [f'(\tau) f'(\tau')]^{1-2\Delta} D(f(\tau), f(\tau')) \\
&\Pi(\tau, \tau') \rightarrow [f'(\tau) f'(\tau')]^{2\Delta} \Pi(f(\tau), f(\tau'))
\end{align}
where $f$ and $g$ are arbitrary functions and 
\be
\Delta = \frac{1-x}{2}
\ee
is a scaling exponent. As discussed in~\cite{Sachdev:2015efa}, $f$ corresponds to the reparametrization (conformal) symmetry, 
and $g$ corresponds to an emergent symmetry of the complex fermions, which is a $U(1)$ gauge symmetry when $|g|=1$.

Using the emergent symmetries, we now obtain the finite-temperature, imaginary-time propagators as well as some other propagators that will be necessary to compute the OTOC, using the method of~\cite{PhysRevD.94.106002} and~\cite{Murugan2017}. Focusing first on the bosons, we make the choice 
\begin{align}
f(\tau) = \tan( \frac{\pi\tau}\beta) 
\end{align}
 in equation \eqref{Dmapping} to obtain
\begin{align}
\label{Dimag}
D(\tau-\tau') = A_D \qty(\frac{\pi}{\beta |\sin \frac{\pi(\tau - \tau')}{\beta}|})^{2x},
\end{align}
which solves the Schwinger-Dyson equations, is translationally invariant, and has the required periodicity in $\beta$. This propagator is defined as $D(\tau-\tau') \equiv \expval{T \phi(\tau) \phi(\tau')}$ where the $\tau$'s are in general complex and, importantly, $T$ orders the fields according to the \textit{real} part of $\tau$ (corresponding to imaginary time) and ignores the imaginary part (corresponding to real time). This means that the retarded propagator $-iD^R(t-t') \equiv \expval{[\phi(it),\phi(it')]}\theta(t-t')$ can be built from \eqref{Dimag} by adding an infinitesimal real part to the arguments to ensure the correct ordering:
\begin{align}
D^R(t-t') = -i[D(\ep + it, it') - D(it, \ep + it')]\theta(t-t') \nonumber\\
= -2 A_D \sin(\pi x) \qty(\frac{\pi}{\beta \sinh \frac{\pi (t-t')}{\beta}})^{2x} \theta(t - t').
\end{align}
We will also need the 
Wightman
propagator which connects two times separated by half the thermal circle:
\begin{align}
D^{lr}(t,t') \equiv \expval{\phi(\beta/2+it)\phi(it')} \nonumber\\
= A_D \qty(\frac{\pi}{\beta\cosh\frac{\pi(t-t')}{\beta}})^{2x}.
\end{align}
Because this propagator is automatically time-ordered, it is simply $D(\beta/2+it,it')$.

Now we do the same for the fermions, applying the transformation \eqref{Gmapping} with the aforementioned choice of $f(\tau)$ and using time translation invariance to set $\tau' = 0$. We have an additional freedom to choose $g(\tau)$ in \eqref{Gmapping}. 
The finite-temperature Green's function is then, with $g(\tau)$ still undetermined,
\begin{align}
\label{GfiniteT}
&G(\tau) = -A_G g(\tau)\sgn[\tan (\pi\tau/\beta)] \nonumber\\
&\times\cos(\frac{\pi x}{2} +\sgn[\tan (\pi\tau/\beta)]\arctan\alpha) 
\qty(\frac{\pi}{\beta|\sin\frac{\pi\tau}{\beta}|})^{1-x}.
\end{align}
To determine $g(\tau)$, we use the short-time information of the Green's function which enforces the antiperiodicity condition $G(\tau-\beta) = -G(\tau)$.
The result is the constraint
\begin{align}
\frac{g(\tau-\beta)}{g(\tau)} = \frac{\cos(\frac{\pi x}{2} + \arctan \alpha)}{\cos(\frac{\pi x}{2} - \arctan\alpha)}
\end{align}
which is solved by 
\begin{align}
\label{normalizedg}
g(\tau) = \(\frac{\cos(\frac{\pi x}{2} + \arctan \alpha)}{\cos(\frac{\pi x}{2} - \arctan\alpha)}\)^{-\tau/\beta},
\end{align}
normalized such that $g(\tau)=1$ at $T=0$.

The propagator \eqref{GfiniteT}, with $g(\tau)$ given by \eqref{normalizedg}, is $G(\tau-\tau')\equiv\expval{Tc(\tau)c\da(\tau')}$, from which we can build the retarded propagator
\begin{align}
G^R(t-t') \equiv &\expval{\{c(it),c\da(it')\}}\theta(t-t')  \nonumber\\
=&[G(\ep+it,it')-G(it,\ep+it')]\theta(t-t') \nonumber\\
=& -A_G e^{i\arctan\alpha}\sin(\pi x)
g(i(t-t'))
\nonumber\\
&\times\qty(\frac{\pi}{\beta \sinh\frac{\pi(t-t')}{\beta}})^{1-x}
\theta(t-t')
\end{align}
and (note the argument -- these propagators are defined so that the later time is always called $t$)
\begin{align}
G^A(t-t') \equiv \expval{\{c(it'),c\da(it)\}}\theta(t-t') \nonumber\\
= [G^R(t-t')]^*.
\end{align}
 The Wightman propagators are given by
\begin{align}
&G^r(t-t') \equiv \expval{c(\beta/2+it) c\da(it')}
= G(\beta/2+it,it') \nonumber\\
=& -A_G \sqrt{\cos(\frac{\pi x}{2} - \arctan\alpha)\cos(\frac{\pi x}{2} + \arctan\alpha)} \nonumber\\
&\times g(i(t-t'))
\qty(\frac{\pi}{\beta\cosh\frac{\pi(t-t')}{\beta}})^{1-x}
\end{align}
and 
\begin{align}
G^l(t-t') \equiv \expval{c(it')c\da(\beta/2+it)} = [G^r(t-t')]^*.
\end{align}

\subsection{Insulating state}
At zero temperature and in the weak coupling limit $\omega_0 \ll m_0$, the system has a (meta)stable insulating phase for $\mu > \omega_F/2$~\cite{Wang:2020dtj}. In this phase, the Green's functions are simply those of free bosons (with the mass unchanged) and free fermions with a renormalized chemical potential 
\begin{align}
\tilde{\mu} \equiv \mu - \frac{\omega_F}2 > 0.
\end{align}
Fourier transforming to imaginary time yields
\begin{align}
G(\tau) = e^{-\tilde{\mu}\tau}\theta(\tau), \quad\quad
D(\tau) = \frac{e^{-m_0|\tau|}}{2 m_0}.
\end{align}
Because of the gapped nature of the phase, the Green's functions are approximately independent of temperature as long as $T\ll \tilde \mu, m_0$. The retarded and advanced Green's functions are constructed as before (this can also be done from the frequency domain with the usual analytic continuation $i\omega \rightarrow \omega \pm i\delta$):
\begin{align}
\label{insulatingretadv}
G^{R,A}(t-t') = e^{\mp i\tilde{\mu}(t-t')}\theta(t-t'), \nonumber\\
D^{R}(t-t') = \frac{\sin[m_0 (t-t')]}{m_0} \theta(t-t').
\end{align}
Finally, we obtain the right- and left-pointing Wightman propagators
\begin{align}
\label{insulatingrightleft}
G^{r,l}(t-t') = e^{-\tilde{\mu}[\frac{\beta}{2} \pm i(t-t')]}, \nonumber\\
D^{r,l}(t-t') = \frac{1}{2m_0} e^{-m_0[\frac{\beta}{2} \pm i(t-t')]}.
\end{align}
However, the Wightman propagators for the bosons contain a factor of $\exp(-m_0\beta/2)$ and are exponentially suppressed, since we assume $m_0 \gg T$. (The fermion propagators are not similarly suppressed because we make no assumption about the ratio $\tilde{\mu}/T$.) This means that we can approximate $D^{r,l} \approx 0$, and we will entirely neglect the diagrams in the ladder series with this type of rung.

\section{Solution of the ladder equation}
\label{sec:3}
For the Yukawa-SYK model, the OTOC in Eq.~\eqref{eq:1} is precisely defined as
\be
\label{eq:fermionOTOC}
F_c(t_1,t_2) \! = \!\sum_{ij}
\frac{\Tr}{M^2}[\rho^\frac{1}{2} \{c_i(t_1), c_j\da(0)\} \rho^\frac{1}{2} \{c_i(t_2), c_j\da(0)\}\da],
\ee

\begin{figure}
\includegraphics[width=0.5\columnwidth]{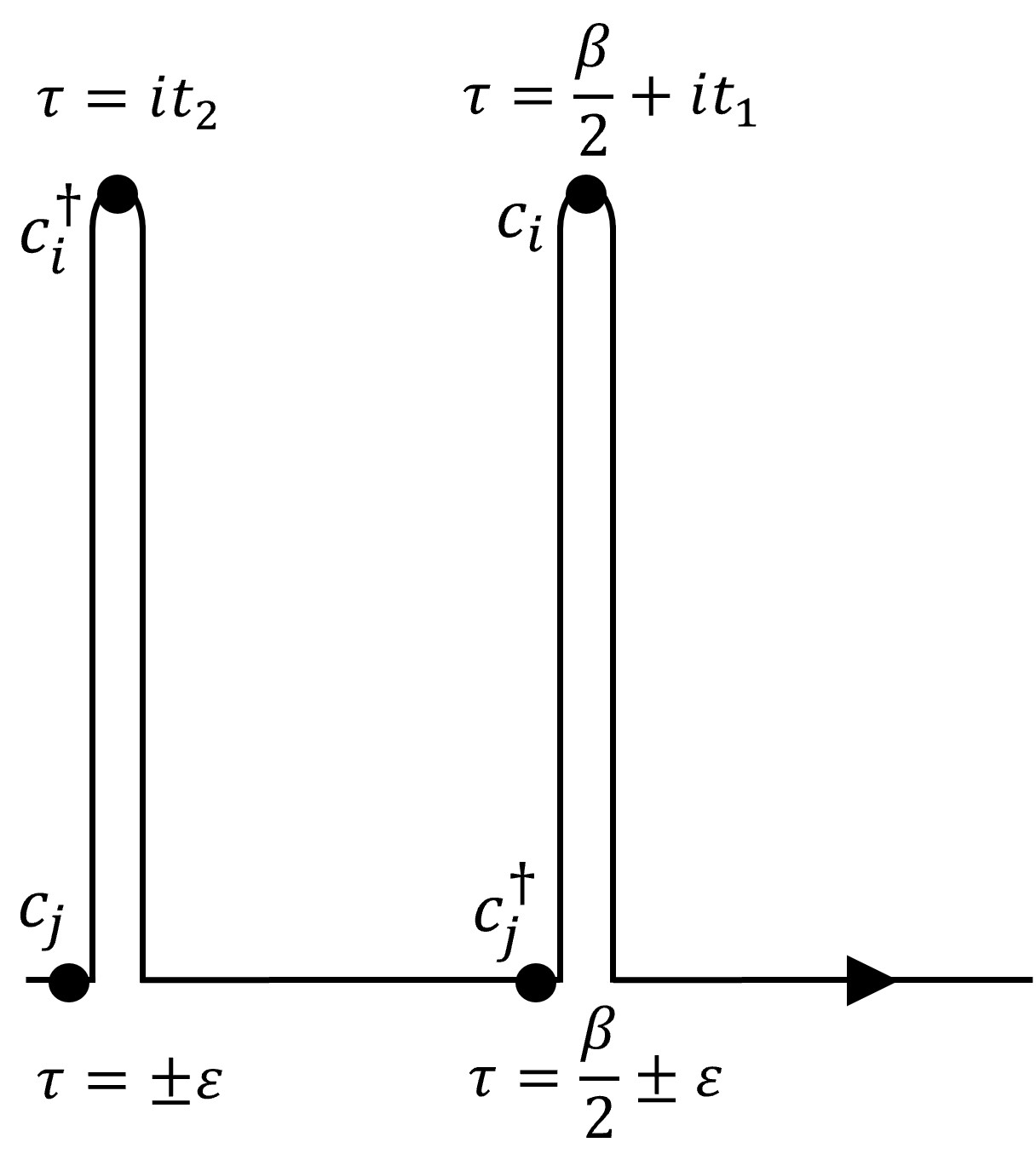}
\caption{The Keldysh contour used for the expansion of the OTOC, with two real-time folds (note that real-time evolution corresponds to the imaginary part of $\tau$). The real part of $\tau$ has period $\beta$; the ends of the contour should be identified. The arrow indicates the direction of contour-ordering. One ordering of the operators, from the expansion of the anticommutators, is depicted.}
\label{contour}
\end{figure}

We evaluate the OTOC by the usual technique which is described in~\cite{Murugan2017} (see also~\cite{Stanford:2015owe,Marcus2018ANC}) via a path integral defined on a complex time (Keldysh) contour, which encircles the compactified imaginary time direction and also has two real time folds (``rails'') at $\tau = 0$ and $\tau = \beta/2$ (see Fig.~\ref{contour}). Two operators are placed at the two ends ($\Im \tau=0$ and $\Im \tau=t_{1,2}$) of each rail. Since imaginary time is strictly increasing along this contour, contour-ordering is the same as $\tau$-ordering and we can use the $\tau$-ordered Green's functions of the previous section. The four possible orderings of the operators in the double-anticommutator are accounted for by displacing the operators at $t=0$ infinitesimally forward or backward in the imaginary time direction so that they lie just after or just before the real time folds; summing over the four different contour-orderings yields the desired correlator. 

At zeroth order in the interaction, each of the orderings can be computed by contracting the operators using Wick's theorem in the path integral along the complex contour. The terms with contractions between operators across different rails cancel after summing over different orderings, and the survivors conspire to produce a product of the (free) retarded and advanced Green's functions.

\begin{figure}
\includegraphics[width=0.9\columnwidth]{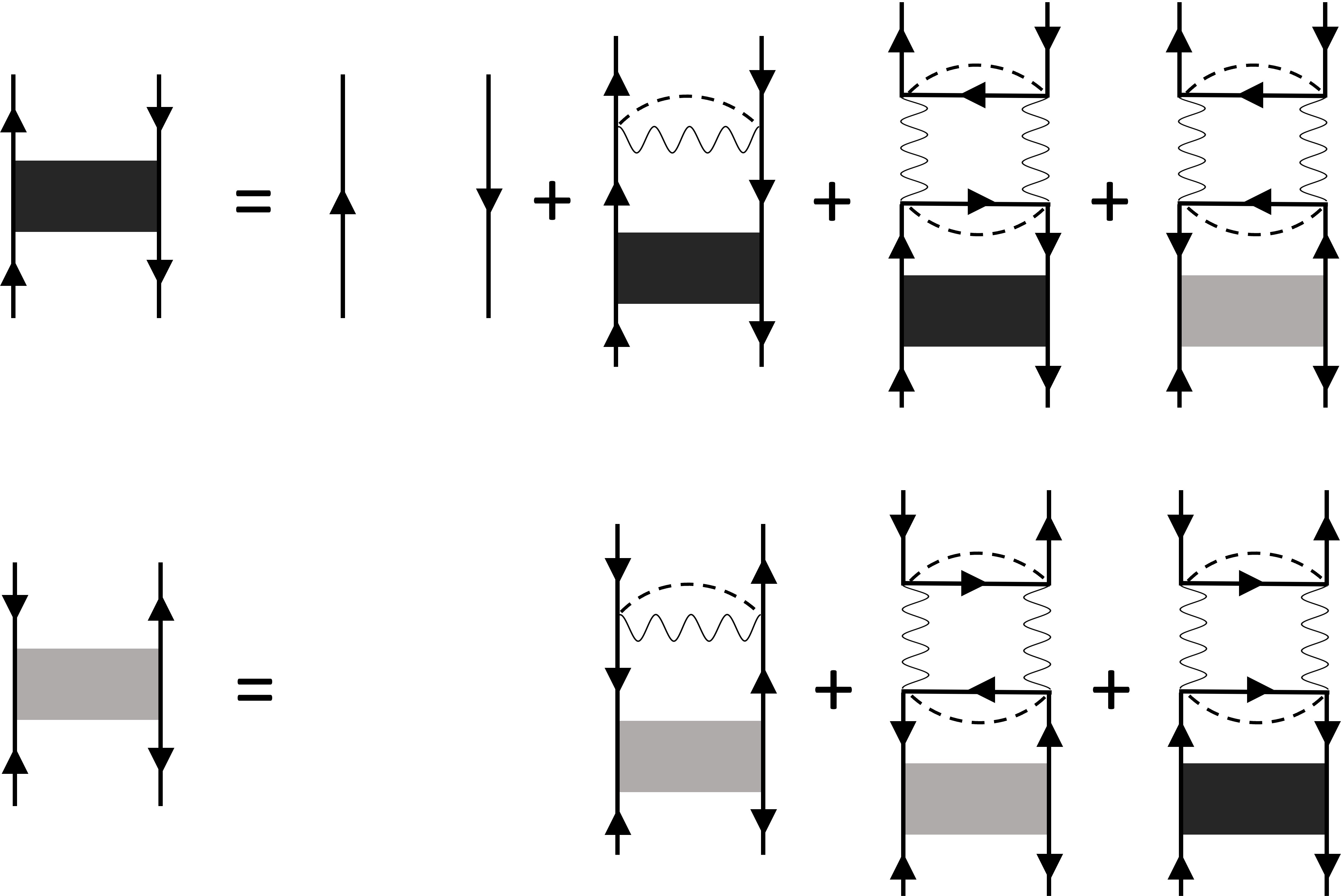}
\caption{{Feynman diagrams for the fermionic OTOC to leading order in $1/N$ and $1/M$. All propagators are fully renormalized by self-energy contributions to leading order in $1/N$ and $1/M$}. The disorder average involves averaging products of independent Gaussian variables, which can be reduced by Isserlis's theorem to a sum of products of averages of all pairings of the variables. Hence the disorder average, indicated by dashed lines, acts like another field with a constant propagator which requires the fermion and boson flavors at one end to match those at the other end. Diagrams with crossed rungs, and other types of pairings in the disorder average, are suppressed in the large $M,N$ limit.}
\label{fermionladder}
\end{figure}

In the interaction picture, we expand the interaction-induced time evolution operator perturbatively, inserting copies of the interaction vertex along the real time folds and using Wick's theorem. Insertions along the imaginary time axis, on the other hand, cancel upon summing over orderings. When operators on the same rail are contracted, the resulting contributions are just part of the self-energy, and these are accounted for by using the fully dressed propagators rather than the non-interacting ones. On the contrary, the contractions within interaction vertices between opposite rails are not captured by the self-energy. Taking into account these diagrams (with insertions on both sides of each rail), to leading order in $M$ and $N$, we find that the OTOC can be expressed as a series of uncrossed ladder diagrams, which can be written recursively as in Fig.~\ref{fermionladder}. {The fact that only ladder diagrams need to be included at leading order in large-$N$ is a common theme in related models and has been shown, for example, for the SYK model~\cite{PhysRevD.94.106002} and for a weak coupling $\phi^4$ theory~\cite{Stanford:2015owe}. For the details of the large-$M,N$ analysis for the current model, see Appendix~\ref{appendix:A}.} According to a modified Feynman rule (cf.~\cite{Stanford:2015owe}), the vertical lines are retarded (advanced) propagators and the horizontal lines are Wightman propagators. For completeness, in Appendix~\ref{appendix:B}, we prove the Feynman rule using the lowest order diagram of the ladder series. 

The goal is to compute the correlator on the LHS of the upper equation in Fig.~\ref{fermionladder}. As is indicated by the last diagram on the RHS, its full expression involves a second correlator, defined as
 \be
F_{\bar c} = \frac{1}{M^2}\sum_{ij}\Tr[\rho^\frac{1}{2} \{c_i\da(t_1), c_j\da(0)\} \rho^\frac{1}{2} \{c_i\da(t_2), c_j\da(0)\}\da],
\ee 
given by the lower equation, to account for the two possible directions of fermion loops (a complication that does not arise in models with Majorana fermions). Even though $F_{\bar c}=0$ for free fermions, it is generated by interaction effects at second order in perturbation theory. To see this, one can plug the first term on the right hand side of the upper equation in Fig.~\ref{fermionladder} to the third term of the right hand side of the lower equation. More detail on how this diagram follows from the Feynman rules can be found in Appendix~\ref{appendix:C}. In turn, $F_{\bar c}$ contributes to $F_c$ via the last diagram on the RHS of the upper equation in Fig.~\ref{fermionladder}. Following the Feynman rules in a similar way, we show the leading order contribution at fourth order in Appendix~\ref{appendix:C}.

As we shall see, including these diagrams is important for getting the correct Lyapunov exponent. We note that there is some variation in the literature as to whether this type of diagram is included, which we leave for future investigations. In agreement with Refs.~\cite{PhysRevB.103.L081113} and~\cite{2020JHEP...07..055S}, we find that for complex fermions such diagrams are indeed generated.

We express the ladder equation for the OTOC in terms of integral kernels
\begin{align}
\label{allkernels}
K_{cc} &= \omega_0^3 G^R(t_{13})G^A(t_{24})D^{lr}(t_{43}) \nonumber\\
K_{\phi c} &= \omega_0^3 D^R(t_{13}) D^R(t_{24}) G^r(t_{43}) \nonumber\\
K_{c \phi} &= \frac{M}{N} \omega_0^3 G^R(t_{13}) G^A(t_{24}) G^l(t_{43}) \nonumber\\
K_{\bar{c}\bar{c}} &= \omega_0^3 G^A(t_{13})G^R(t_{24})D^{lr}(t_{43}) \nonumber\\
K_{\phi \bar{c}} &= \omega_0^3 D^R(t_{13}) D^R(t_{24}) G^l(t_{43}) \nonumber\\
K_{\bar{c} \phi} &= \frac{M}{N} \omega_0^3 G^A(t_{13}) G^R(t_{24}) G^r(t_{43})
\end{align}
where $t_{ij} \equiv t_i - t_j$. The notation $K_{\phi c}$, for example, indicates that it corresponds to portions of the diagrams which create bosons at later times out of incoming fermions. The $\bar{c}$ refers to fermion lines with arrows that are reversed compared to the OTOC (that is, an advanced propagator on the left rail and a retarded one on the right). The kernels involving $\bar{c}$ are simply the complex conjugates of the the corresponding kernels with $c$. Denoting the fermion OTOC by $F_{c}(t_1,t_2)$, the ladder equation is then
\begin{align}
\label{ladder}
F_{c} =& \frac{1}{M}G^R(t_1)G^A(t_2) \nonumber\\
&+K_{cc} \star F_{c} + K_{c\phi} \star K_{\phi c} \star F_{c} 
+ K_{c\phi} \star K_{\phi \bar{c}} \star F_{\bar c} \nonumber\\
F_{\bar c} =& K_{\bar{c}\bar{c}} \star F_{\bar c} + K_{\bar{c}\phi} \star K_{\phi\bar{c}} \star F_{\bar c}
+ K_{\bar{c}\phi} \star K_{\phi c} \star F_{c}
\end{align}
The $\star$ denotes an integral convolution, that is,
\begin{align}
(K \star F)(t_1,t_2) \equiv \int_{-\infty}^{\infty} \dd{t_3}\dd{t_4} K(t_1,t_2,t_3,t_4)F(t_3,t_4).
\end{align}
The inhomogeneous first term $G^R(t_1)G^A(t_2)/M$ in the series 
does not increase exponentially like $F_c$
and quickly becomes negligible. The inhomogeneous term acts as a source and ensures that the exponential growth is suppressed by $1/M$, i.e., $F_c \propto e^{\lambda_L t}/M$. The value of $\lambda_L$ can be determined by dropping the inhomogeneous term. After doing this, the two equations in Fig.~\ref{fermionladder} can be solved by $F_c=F_{\bar c}^*$, so that one only needs to solve one of them.

The strategy for solving this eigenfunction problem is to find functions $f_c(t_1,t_2)$ (possibly complex) and $f_\phi(t_1,t_2)$ (real) which satisfy
\begin{align}
\label{kerneleqs}
K_{cc} \star f_c = k_{cc} f_c, ~~&K_{\bar{c}\bar{c}} \star f_c^* = k_{cc} f_c^*  \nonumber\\ 
K_{\phi c} \star f_c = k_{\phi c} f_\phi,~~&K_{\phi\bar{c}} \star f_c^*= k_{\phi c} f_\phi  \nonumber\\
K_{c \phi} \star f_\phi = k_{c \phi} f_c,~~&K_{\bar{c}\phi} \star f_\phi = k_{c\phi} f_c^*,
\end{align}
where $k_{cc}, k_{\phi c}, k_{c\phi}$ are real constants. 
If we can find such functions, then we can choose $F_{c} \sim f_c/M$ and $F_{\bar c} \sim f_c^*/M$ in~\eqref{ladder}. Both equations reduce to the same algebraic requirement that 
\be
k_{cc} + 2k_{c\phi}k_{\phi c} = 1,
\label{eq:42}
\ee
where the factor of 2 comes from considering both $F_{c}$ and $F_{\bar c}$ on the right hand side of the upper equation of Fig.~\ref{fermionladder}.
\subsection{Non-Fermi liquid state}

We begin by writing out $K_{cc}$ in full:
\begin{widetext}
\begin{align}
K_{cc} = 
C_{cc} e^{-it_{13}L/\beta}\qty(\frac{\pi}{\beta\sinh\frac{\pi t_{13}}{\beta}})^{1-x} \theta(t_{13})e^{+it_{24}L/\beta}\qty(\frac{\pi}{\beta\sinh\frac{\pi t_{24}}{\beta}})^{1-x} \theta(t_{24}) 
\qty(\frac{\pi}{\beta\cosh\frac{\pi t_{34}}{\beta}})^{2x}
\end{align}
where
\be
C_{cc} \equiv \omega_0^3 A_G^2 A_D \sin^2(\pi x),\textrm{ and }
L=\log(\frac{\cos(\frac{\pi x}{2} + \arctan \alpha)}{\cos(\frac{\pi x}{2} - \arctan\alpha)}).
\ee
We first make the change of variable $t_i \equiv \frac{\beta}{2\pi}\phi_i$ and use the step functions to restrict the region of integration. The first equation in \eqref{kerneleqs} becomes
\begin{align}
\int_{-\infty}^{\phi_1} \dd{\phi_3} \int_{-\infty}^{\phi_2} \dd{\phi_4} 
C_{cc} e^{-i\phi_{13}L/2\pi} e^{+i\phi_{24}L/2\pi} \qty(\frac{1}{2\sinh \frac{\phi_{13}}{2}})^{1-x} \qty(\frac{1}{2\sinh \frac{\phi_{24}}{2}})^{1-x} 
\qty(\frac{1}{2\cosh \frac{\phi_{34}}{2}})^{2x}
f_c(\phi_3,\phi_4) = k_{cc} f_c(\phi_1,\phi_2)
\end{align}
Now we perform another change of variable: $z_{1,3} \equiv e^{-\phi_{1,3}}, z_{2,4} \equiv -e^{-\phi_{2,4}}$ (note the minus sign). The kernel equation becomes
\begin{align}
C_{cc} \int_{z_1}^{\infty} \frac{\dd{z_3}}{|z_3|} \int_{-\infty}^{z_2} \frac{\dd{z_4}}{|z_4|}
\qty|\frac{z_1 z_4}{z_2 z_3}|^{\frac{iL}{2\pi}}
\frac{|z_1 z_3|^{\frac{1}{2}(1-x)}}{|z_{13}|^{1-x}}
\frac{|z_2 z_4|^{\frac{1}{2}(1-x)}}{|z_{24}|^{1-x}}
\frac{|z_3 z_4|^x}{|z_{34}|^{2x}} f_{c}(z_3,z_4) = k_{cc} f_{c}(z_1,z_2),
\label{eq:kernel}
\end{align}
\end{widetext}
where $z_{ij}\equiv z_i-z_j$.
We solve this by making the ansatz
\begin{align}
\label{fc}
f_c(z_1,z_2) = \qty|\frac{z_1}{z_2}|^\frac{iL}{2\pi}
\frac{|z_1 z_2|^{\frac{1}{2}(1-x)}}{|z_{12}|^{1-x-h}}
\end{align}
which will make the integrand only a function of the difference between the $z$'s.
With this choice the integral in~\eqref{eq:kernel} becomes
\begin{align}
C_{cc}\qty( \int_{z_1}^{\infty} \!\!\!\!\! \dd{z_3} \int_{-\infty}^{z_2} \!\!\!\!\! \dd{z_4}
\frac{|z_{12}|^{1-x-h}}{|z_{13}|^{1-x}|z_{24}|^{1-x}|z_{34}|^{1+x-h}} )
f_c(z_1,z_2).
\end{align}
Here $h$ is a parameter that can be later tuned to satisfy~\eqref{eq:42}. As we shall see, it is directly related to $\lambda_L$. The expression in parentheses, as promised, is simply a constant, despite appearing to be a function of $z_1$ and $z_2$. It is invariant under the affine transformation $z_i \rightarrow a z_i + b$; in particular the integration variables can be shifted and scaled to set $z_1 = 1, z_2 = 0$ and the integral can be evaluated:
\begin{align}
k_{cc} =& 
 C_{cc} \int_{1}^{\infty} \dd{z_3} \int_{-\infty}^{0} \dd{z_4}
\frac{1}{|1-z_3|^{1-x}|z_4|^{1-x}|z_{34}|^{1+x-h}} \nonumber\\
=& C_{cc} \frac{\Gamma^2(x)\Gamma(1-x-h)}{\Gamma(1+x-h)}.
\end{align}

Now we repeat this procedure for the second equation of \eqref{kerneleqs}, feeding in our expression for $f_c$. Upon making the additional ansatz
\begin{align}
\label{fphi}
f_\phi(z_1,z_2) = \frac{|z_1 z_2|^x}{|z_{12}|^{2x-h}},
\end{align}
again to make the integrand dependent only on the $z_{ij}$'s,
we find
\begin{align}
k_{\phi c} &= 
 C_{\phi c} \int_{1}^{\infty} \dd{z_3} \int_{-\infty}^{0} \dd{z_4}
\frac{1}{|1-z_3|^{2x}|z_4|^{2x}|z_{34}|^{2-2x-h}} \nonumber\\
&= C_{\phi c} \frac{\Gamma^2(1-2x)\Gamma(2x-h)}{\Gamma(2-2x-h)}
\end{align}
where 
\begin{align}
C_{\phi c} \equiv -4 \omega_0^3 A_G A_D^2 \sin^2(\pi x) (2\pi/\beta)^{3x-1}\nonumber\\
\times \sqrt{\cos(\frac{\pi x}{2} - \arctan\alpha)\cos(\frac{\pi x}{2} + \arctan\alpha)}.
\end{align}

The final two equations of \eqref{kerneleqs} are compatible with the same choice of $f_c$ and $f_\phi$, and the result is
\begin{align}
k_{c\phi} &=
C_{c\phi} \int_{1}^{\infty} \dd{z_3} \int_{-\infty}^{0} \dd{z_4}
\frac{1}{|1-z_3|^{1-x}|z_4|^{1-x}|z_{34}|^{1+x-h}} \nonumber\\
&= C_{c \phi} \frac{\Gamma^2(x)\Gamma(1-x-h)}{\Gamma(1+x-h)}
\end{align}
where
\begin{align}
C_{c \phi} \equiv -\frac{M}{N} \omega_0^3 A_G^3 \sin^2(\pi x) (2\pi/\beta)^{1-3x} \nonumber\\
\times \sqrt{\cos(\frac{\pi x}{2} - \arctan\alpha)\cos(\frac{\pi x}{2} + \arctan\alpha)}.
\end{align}

With these results, Eq.~\eqref{eq:42} reduces to an algebraic equation for $h$,
\begin{align}
\label{eqforh}
&\frac{\Gamma^2(x)\Gamma(1-x-h)}{\Gamma(1+x-h)}\nonumber\\
&\times\qty[C_{cc} + 2C_{c\phi}C_{\phi c} \frac{\Gamma^2(1-2x)\Gamma(2x-h)}{\Gamma(2-2x-h)}]
=1.
\end{align}
The constants $C_{cc},C_{c\phi},C_{\phi c}$ involve a complicated mix of $\alpha, \beta,  \omega_0, m_0, \omega_f, \frac{n_s M}{N}, \mbox{ and } x$. 
We can use the relations~\eqref{parameterIdentities}, 
along with the gamma function reflection identity and the fact $\cos(\pi x/2-\arctan\alpha)\cos(\pi x/2 + \arctan\alpha) = \frac{1}{2}(\cos(\pi x) + \frac{1-\alpha^2}{1+\alpha^2})$, to write \eqref{eqforh} purely in terms of $x$:
\begin{align}
\label{constraintonh}
&1 = \frac{-1}{\Gamma(-x)\Gamma(x)}
\frac{\Gamma^2(x) \Gamma(1-x-h)}{\Gamma(1+x-h)} \nonumber\\
&\times \qty[1 
+ 2(1-2x)\frac{1}{\Gamma(2x)\Gamma(1-2x)}
\frac{\Gamma^2(1-2x)\Gamma(2x-h)}{\Gamma(2-2x-h)}].
\end{align}
{Remarkably, this is exactly solved for all $x$ by 
\be
h = -1.
\ee 
This can be verified analytically by making repeated use of the identity $\Gamma(z+1)=z\Gamma(z)$.}
Transforming back to the original time variables, we have
\begin{align}
f_c(t_1,t_2) = e^{\frac{-iL}{\beta}(t_1-t_2)}
\frac{e^{-h\frac{2\pi}{\beta}\qty(\frac{t_1+t_2}{2})}}
{(2\cosh{\frac{2\pi}{\beta}\qty(\frac{t_1-t_2}{2})})^{1-x-h}}.
\end{align}
The OTOC grows in the same way as $f_c(t,t)$ at long times, and the Lyapunov exponent can be read off from this expression:
\be
\lambda_L = \frac{-2\pi}{\beta}h = {2\pi}T.
\ee
 Notice that the only effect of the chemical potential, which enters through $L$, is to introduce an oscillatory factor in $f_c$ which does not affect the OTOC or $\lambda_L$. This means that in any regime where the conformal limit is valid, the spectral asymmetry does not disturb the OTOC from its form at half-filling, which corresponds to $\mu = \alpha = L = 0$. Furthermore, $\lambda_L$ is independent of the ratio $M/N$.


\begin{figure}
\includegraphics[width=0.7\columnwidth]{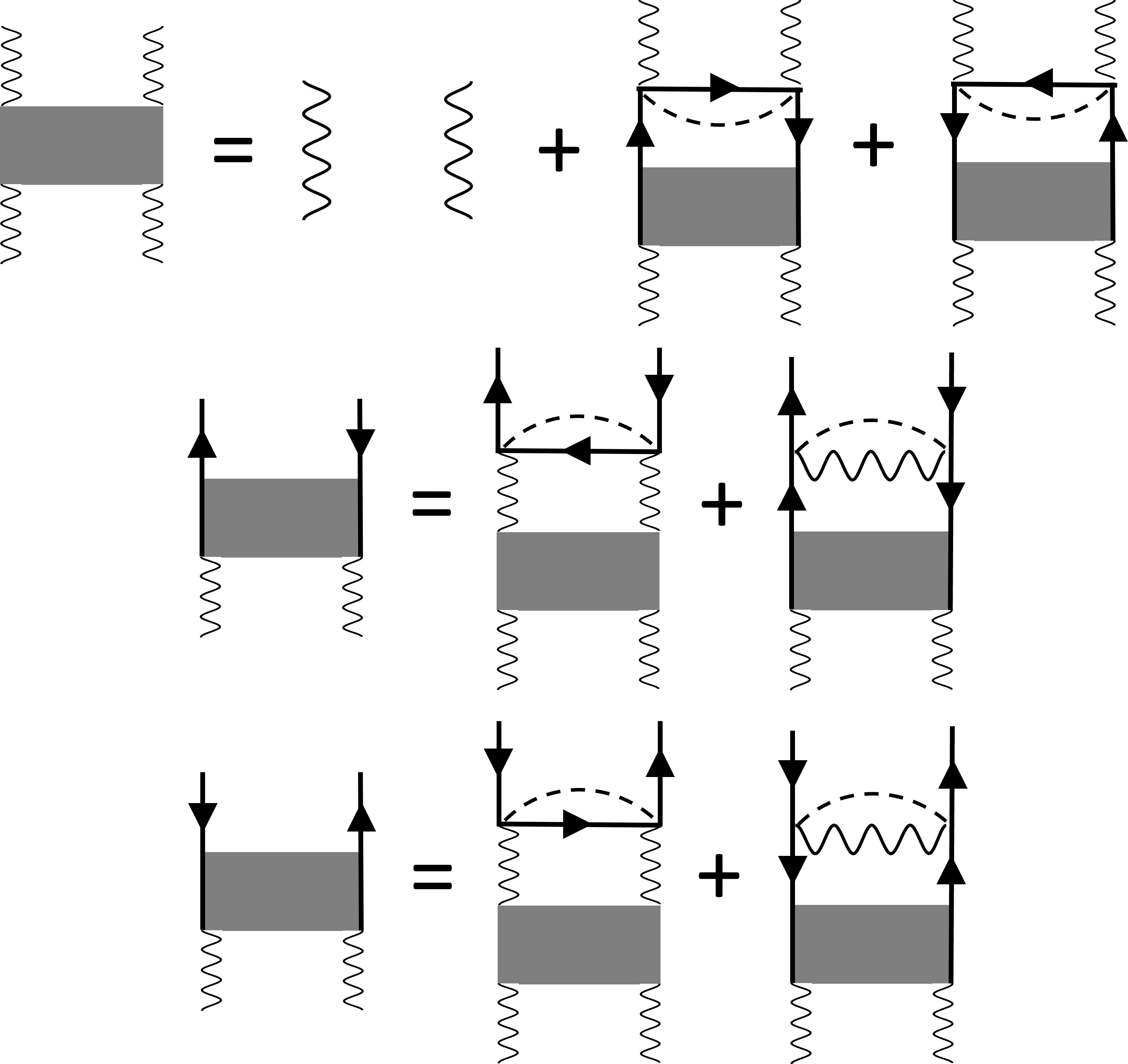}
\caption{Feynman diagrams for the bosonic OTOC. Notation is the same as in Fig.~\ref{fermionladder}.}
\label{bosonladderfig}
\end{figure}
{Interestingly, by following the argument in~\cite{Marcus2018ANC}, we can conclude that the bosons are maximally chaotic as well with little additional computation, since the boson ladder equation involves the same kernels as for the fermions.} The boson correlator $F_{\phi}(t_1,t_2)$ is defined as
\be
F_{\phi}(t_1,t_2) \! = \!\!\sum_{\alpha\beta} \!
\frac{\Tr}{N^2}[\rho^\frac{1}{2} [\phi_\alpha(t_1), \phi_\beta(0)] \rho^\frac{1}{2} [\phi_\alpha(t_2), \phi_\beta(0)]\da],
\ee
and can be expressed diagrammatically via the coupled equations of Fig.~\ref{bosonladderfig} by introducing the correlators $F_{c \phi}$ and $F_{\bar{c} \phi}$, which involve the processes that convert two incoming bosons to two outgoing fermions. Algebraically, the equations read 
\begin{align}
\label{bosonladdereqs}
F_{\phi} = F_{\phi}^0 + K_{\phi c} \star F_{c \phi} + K_{\phi \bar{c}} \star F_{\bar{c}\phi} \nonumber\\
F_{c \phi} = K_{cc} \star F_{c\phi} + K_{c\phi} \star F_{\phi} \nonumber\\
F_{\bar{c} \phi} = K_{\bar{c}\bar{c}} \star F_{\bar{c}\phi} + K_{\bar{c}\phi} \star F_{\phi},
\end{align}
where $F_{c\phi}$, $F_{\bar{c}\phi}$, and etc., represented diagrammatically in Fig.~\ref{bosonladderfig}, are ``mixed" OTOC's defined as, e.g.,
\be
F_{c\phi}(t_1,t_2) \! = \!\! \sum_{i \alpha} \!
\frac{\Tr}{MN} \qty [\rho^\frac{1}{2} [c_i(t_1), \phi_\alpha(0)] \rho^\frac{1}{2} [c_i(t_2), \phi_\alpha(0)]\da].
\ee
We neglect the inhomogeneous term in \eqref{bosonladdereqs} as usual and substitute the first equation into the second and third to eliminate $F_{\phi}$. We see that $F_{c\phi}$ and $F_{\bar{c}\phi}$ obey precisely the same ladder equation as $F_{c}$ and $F_{\bar c}$ respectively (cf.~Eq.~\eqref{ladder}), and thus they too are proportional to $f_c$ and $f_c^*$ with the same exponent $h = -1$. Hence the mixed OTOCs such as $F_{c\phi}$ also have the same Lyapunov exponent. 

Substituting these solutions back into the first equation of~\eqref{bosonladdereqs} and using the properties of the kernels~\eqref{kerneleqs}, we find that the boson OTOC $F_{\phi}$ is proportional to $f_\phi$, which in the time domain is
\begin{align}
f_\phi(t_1,t_2) =
\frac{e^{-h\frac{2\pi}{\beta}\qty(\frac{t_1+t_2}{2})}}
{(2\cosh{\frac{2\pi}{\beta}\qty(\frac{t_1-t_2}{2})})^{2x-h}}.
\end{align}
This has exactly the same long-time growth and Lyapunov exponent as the fermion OTOC, only without the oscillation in $t_1-t_2$. 

Intuitively, since the system is strongly interacting among both boson and fermions, the exponential growth of chaos does not depend on which field operators are used to perturb and probe the system.

\subsection{Insulating state}
We repeat the procedure for the (meta)stable insulating state. As mentioned earlier (cf.~Eq.~\eqref{insulatingrightleft}), the boson left-right propagators are exponentially suppressed, so we drop the inhomogeneous terms, containing $K_{cc}$ and $K_{\bar{c}\bar{c}}$, on the right-hand sides of \eqref{ladder}:
\begin{align}
\label{ladderins}
F_{c} = K_{c\phi} \star K_{\phi c} \star F_{c} + K_{c\phi} \star K_{\phi \bar{c}} \star F_{\bar c} \nonumber\\
F_{\bar c} = K_{\bar{c}\phi} \star K_{\phi\bar{c}} \star F_{\bar c} + K_{\bar{c}\phi} \star K_{\phi c} \star F_{c}.
\end{align}
The kernels have the same structure as before, but the Green's functions are now the insulating ones. Again we solve this by finding $f_c$ and $f_\phi$ satisfying~\eqref{kerneleqs}.
\begin{align}
\label{kerneleqs3}
K_{\phi c} \star f_c = k_{\phi c} f_\phi \nonumber\\
K_{c \phi} \star f_\phi = k_{c \phi} f_c
\end{align}
 Then the choice $F_{c} = f_c, F_{\bar c} = f_c^*$ solves the insulating ladder equation when 
 \be
 1 = 2k_{c\phi}k_{\phi c}.
 \ee

By the same argument as in the previous subsection, the boson OTOC $F_\phi$ will again be proportional to the function $f_\phi$. Since we will see that $f_c$ and $f_\phi$ both have the same exponential growth, we conclude that the fermions and bosons have the same Lyapunov exponent in the insulating phase. 

We make the ansatz
\begin{align}
\label{insulatingansatzfc}
f_{c}(t_1,t_2) = e^{\frac{\lambda_L}{2}(t_1+t_2) - i\gamma(t_1-t_2)}
\end{align}
where $\lambda_L$ and $\gamma$ are real, and $\lambda_L$ is of course the Lyapunov exponent. Since we obtain the OTOC, which must be real, by setting $t_1=t_2$, $\gamma$ will drop out of the final result, but it is nonetheless necessary to keep it as a parameter.  As will be shown, in the absence of the inhomogeneous term, $\lambda_L$ and $\gamma$ parametrize a whole family of functions with the properties~\eqref{kerneleqs}, and the requirement $1 = 2k_{c\phi}k_{\phi c}$ constrains the solutions for $\lambda_L$ and $\gamma$. 
The multiplicity of solutions is an artifact of neglecting the inhomogeneous term in the ladder equation (the OTOC should be unique), but physically the ladder equation represents the amplification of $F_{c}^0$ by repeated applications of a kernel, and only the fastest-growing component will survive this procedure. Therefore we will tune $\gamma$ to make $\lambda_L$ as large as possible.

For the first equation in \eqref{kerneleqs3}, we first calculate the action of $K_{\phi c}$ on our ansatz, using 
 Eqs.~\eqref{allkernels},~\eqref{insulatingretadv},~\eqref{insulatingrightleft}, and~\eqref{insulatingansatzfc} the result is
\begin{widetext}
\begin{align}
\label{inskernel1}
K_{\phi c} \star f_c =& \omega_0^3 \int_{-\infty}^{t_1} \dd{t_3} \int_{-\infty}^{t_2} \dd{t_4}
\frac{\sin m_0 t_{13}}{m_0}\frac{\sin m_0 t_{24}}{m_0}
\times e^{-\tilde{\mu}(\beta/2 + i t_{43})}
e^{\frac{\lambda_L}{2}(t_3+t_4) - i \gamma t_{34}} \nonumber\\
=& \frac{\omega_0^3}{m_0^2} e^{-\tilde{\mu}\beta/2} \frac{1}{(2i)^2}
\sum_{\sigma,\sigma'=\pm} \frac{\sigma\sigma'}{(-i\sigma m_0 + i\tilde{\mu} + \lambda_L/2 - i\gamma)(-i\sigma' m_0 - i \tilde{\mu} + \lambda_L/2 + i\gamma)}
e^{-i\tilde{\mu}t_{12}}e^{\frac{\lambda_L}{2}(t_1+t_2)-i\gamma t_{12}}
\end{align}
from which we can read off $k_{\phi c}$, and we find that 
\be
f_\phi(t_1,t_2) = e^{+i\tilde{\mu}t_{12}}e^{\frac{\lambda_L}{2}(t_1+t_2)-i\gamma t_{12}}.
\ee
 Now we act on this with $K_{c\phi}$:
\begin{align}
\label{inskernel2}
K_{c \phi} \star f_\phi = \frac{M}{N} \omega_0^3 \int_{-\infty}^{t_1} \dd{t_3} \int_{-\infty}^{t_2} \dd{t_4}
e^{-i\tilde{\mu}t_{13}} e^{i\tilde{\mu}t_{24}} e^{-\tilde{\mu}(\beta/2 - i t_{43})}
e^{+i\tilde{\mu} t_{34}} e^{\frac{\lambda_L}{2}(t_3+t_4)-i\gamma t_{34}} \nonumber\\
= \frac{M}{N} \omega_0^3 e^{-\tilde{\mu}\beta/2}
\frac{1}{(i\tilde{\mu}+\lambda_L/2-i\gamma)(-i\tilde{\mu}+\lambda_L/2+i\gamma)}
e^{\frac{\lambda_L}{2}(t_1+t_2)-i\gamma t_{12}},
\end{align}
We recover $f_c$ as required, and the prefactor is $k_{c\phi}$. Combining the results of~\eqref{inskernel1} and~\eqref{inskernel2} with the requirement $1 = 2k_{c\phi}k_{\phi c}$, we find that the ansatz solves the ladder equation when
\begin{align}
1 = \frac{2M}{N} \frac{\omega_0^6}{m_0^2} e^{-\tilde{\mu}\beta}
\qty(\frac{1}{2i})^2 \sum_{\sigma \sigma'} \frac{\sigma \sigma'}
{(-i\sigma m_0 + i\tilde{\mu} + \lambda_L/2 - i\gamma)(-i\sigma' m_0 - i\tilde{\mu} + \lambda_L/2 + i\gamma)(i\tilde{\mu} + \lambda_L/2 - i\gamma)(-i\tilde{\mu} + \lambda_L/2 + i\gamma)}.
\label{eq:74}
\end{align}
\end{widetext}
Since we have assumed the boson mass is the largest energy scale in the problem and since $\lambda_L$ cannot exceed $2\pi T$, we can neglect $\tilde{\mu}$ and $\lambda_L$ relative to $m_0$ in the denominator of Eq.~\eqref{eq:74}. We will further assume and then verify that $\gamma$ is similarly negligible. With these assumptions the eigenvalue equation reduces to
\begin{align}
\frac{2M}{N} \omega_F^2 e^{-\tilde{\mu}\beta} = \qty(\frac{\lambda_L}{2})^2 + (\gamma-\tilde{\mu})^2.
\end{align}
Clearly the optimal choice is $\gamma = \tilde{\mu}$, consistent with the assumption $m_0 \gg \gamma$; any other choice drags the Lyapunov exponent down, so
\begin{align}
\lambda_L(\tilde{\mu},T) = \sqrt{\frac{8 M}{N}}\omega_F e^{-\tilde{\mu}/2T} \textrm{, for $T\ll \tilde \mu$}.
\label{eq:ins}
\end{align}
Note that this choice also ensures that the bosonic OTOC is real, just like in the nFL case. We see that for the insulating state $\lambda_L$ is exponentially suppressed, consistent with the understanding that the system is in a trivial gapped state.

Observe that the Lyapunov exponent is an increasing function of temperature and a decreasing function of chemical potential. This is consistent with the intuition that chaos is suppressed as the system goes deeper into the insulating phase, where the filling is close to unity and there is little phase space available for scrambling.

\section{Discussion and conclusion}
\label{sec:4}

In this work, we generalized the Green's functions of the Yukawa-SYK model at weak coupling to finite temperatures and computed the OTOCs and Lyapunov exponents of the nFL state in the conformal limit and the insulating state. The fermions are maximally chaotic in the nFL phase, even away from half-filling. The chemical potential merely gives a phase to the correlator $F_{c}(t_1,t_2)$ which does not affect the OTOC or the Lyapunov exponent. The bosons also saturate the chaos bound, as do the other ``mixed'' correlators described above such as $F_{c\phi}$, etc.

\begin{figure}
\includegraphics[width=\columnwidth]{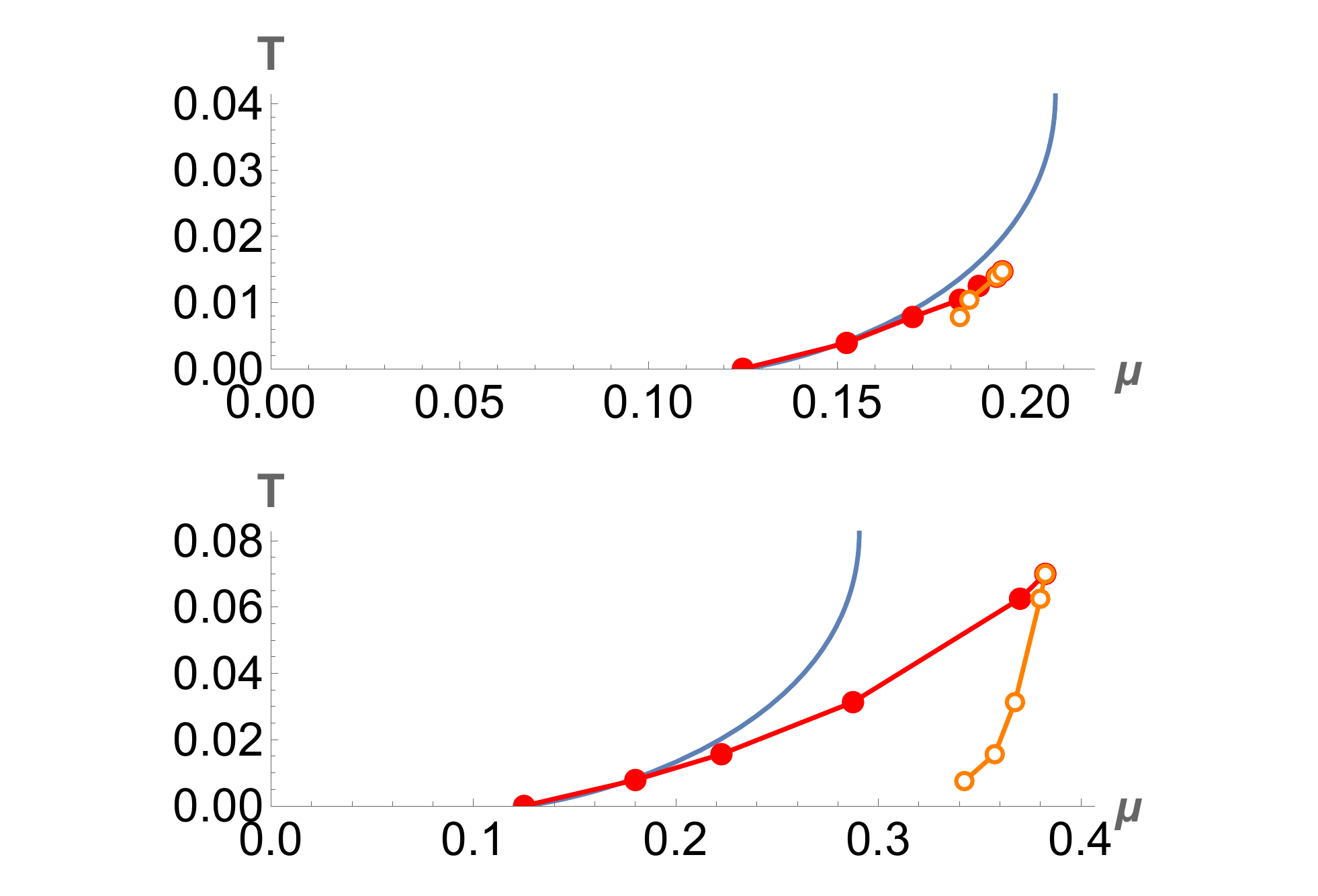}
\caption{The Lyapunov exponent, calculated from the insulating side in the low-$T$ limit, saturates the chaos bound along the solid blue curves. The red curves with data points come from numerical results~\cite{PhysRevB.103.195108} for the leftmost boundary of the hysteresis region. For completeness, the numerical results for the rightmost boundary of the hysteresis region are also shown in orange, though we did not attempt to find this curve analytically. {The $T=0$ data points are analytic results -- see~\cite{Wang:2020dtj}}. The upper and lower panel have $M/N = 1$ and $M/N = 4$ respectively, and both have $\omega_F=1/4$. The agreement is good at low $T$ but, unsurprisingly, fails when $T \sim \omega_F$, noticeable in the lower panel.}
\label{hysteresisboundary}
\end{figure}

For the insulating state, we have obtained an expression for $\lambda_L$ at low temperature in Eq.~\eqref{eq:ins}. As the temperature increases, $\lambda_L$ for the insulating solution increases. As $\lambda_L$ has an upper bound of $2\pi T$, it is interesting to see where in the $(\mu,T)$ plane $\lambda_L$ for the insulating state approaches this value. Heuristically this should correspond to the point when the metastable insulating state becomes unstable.
To determine the boundary in the $(\tilde{\mu},T)$-plane, we set $\lambda_L (T)\sim 2\pi T$ and solve for the critical value $\mu_c$ as a function of temperature in the low-temperature limit:
\begin{align}
\label{maxchaoscurve}
\mu_c(T) = \frac{\omega_F}{2} + 2T \log (\sqrt{\frac{M}{N}}\frac{\omega_F}{T}).
\end{align}
We expect Eq.~\eqref{eq:ins} to be correct for 
$T\ll \tilde \mu\equiv \mu_c  - {\omega_F}/{2} $, so $T\ll \omega_F$. This curve is plotted in blue in Fig.~\ref{hysteresisboundary}.  
To the left of this curve, the exponent violates this bound, so the insulating state cannot be stable, demarcating the leftmost boundary of the hysteresis region (to the right of this curve, we cannot say that the insulating state is energetically favorable to the nFL state, only that it is not forbidden by the chaos bound).

In~\cite{PhysRevB.103.195108}, the nFL and insulating states were placed on a phase diagram as a function of $(\mu, T)$, which we reproduce for the present model in Fig.~\ref{hysteresisboundary}. The data were obtained by iteratively solving the Schwinger-Dyson equations as described therein. (The index structure of the random coupling is slightly different in the model of~\cite{PhysRevB.103.195108}, but the Schwinger-Dyson equations have essentially the same structure. The only change is that $M/N$ in this paper corresponds to $4M/N$ in~\cite{PhysRevB.103.195108}.) The two phases are separated by a first-order transition (not shown), located in the hysteresis region bounded by the red and orange curves, in which both states are local minima of the free energy. On the left branch of the red curve, the insulating state becomes unstable, and on the right, the nFL state becomes unstable.

{The boundary of stability of the insulating state determined by the Lyapunov exponent appears to agree with the numerical results where expected. Because it is difficult to compute the boundary of the hysteresis region numerically by the iterative method described in~\cite{PhysRevB.103.195108}, especially at very low temperatures, and because our equation~\eqref{maxchaoscurve} is a low-temperature approximation, it is difficult to meaningfully quantify this agreement, but nonetheless the connection between the chaos bound and the stability of the phase is plausible.} The rightmost boundary of the hysteresis region, beyond which the nFL state is unstable, will require more work to calculate because our analysis assumes the conformal limit which yields an exponent of $2\pi T$ regardless of $\mu$ and $T$. In reality there should be small corrections to this result, but calculating these will require going beyond the conformal limit. We surmise that the rightmost boundary of the hysteresis region corresponds to the curve where these small corrections switch from negative to positive and violate the chaos bound. In this sense the chaos bound may also be interpreted as a stability bound~\cite{kivelson-2021}. We leave a detailed analysis of this to a future study.
\bigskip

\acknowledgments
We thank A.V.~Chubukov and J.~Schmalian for useful discussions. This work is supported by startup funds at the University of Florida and by NSF under award number DMR-2045781.

\appendix

\section{Large-$M$,$N$ counting}
\label{appendix:A}

\begin{figure}[b]
\includegraphics[width=0.7\columnwidth]{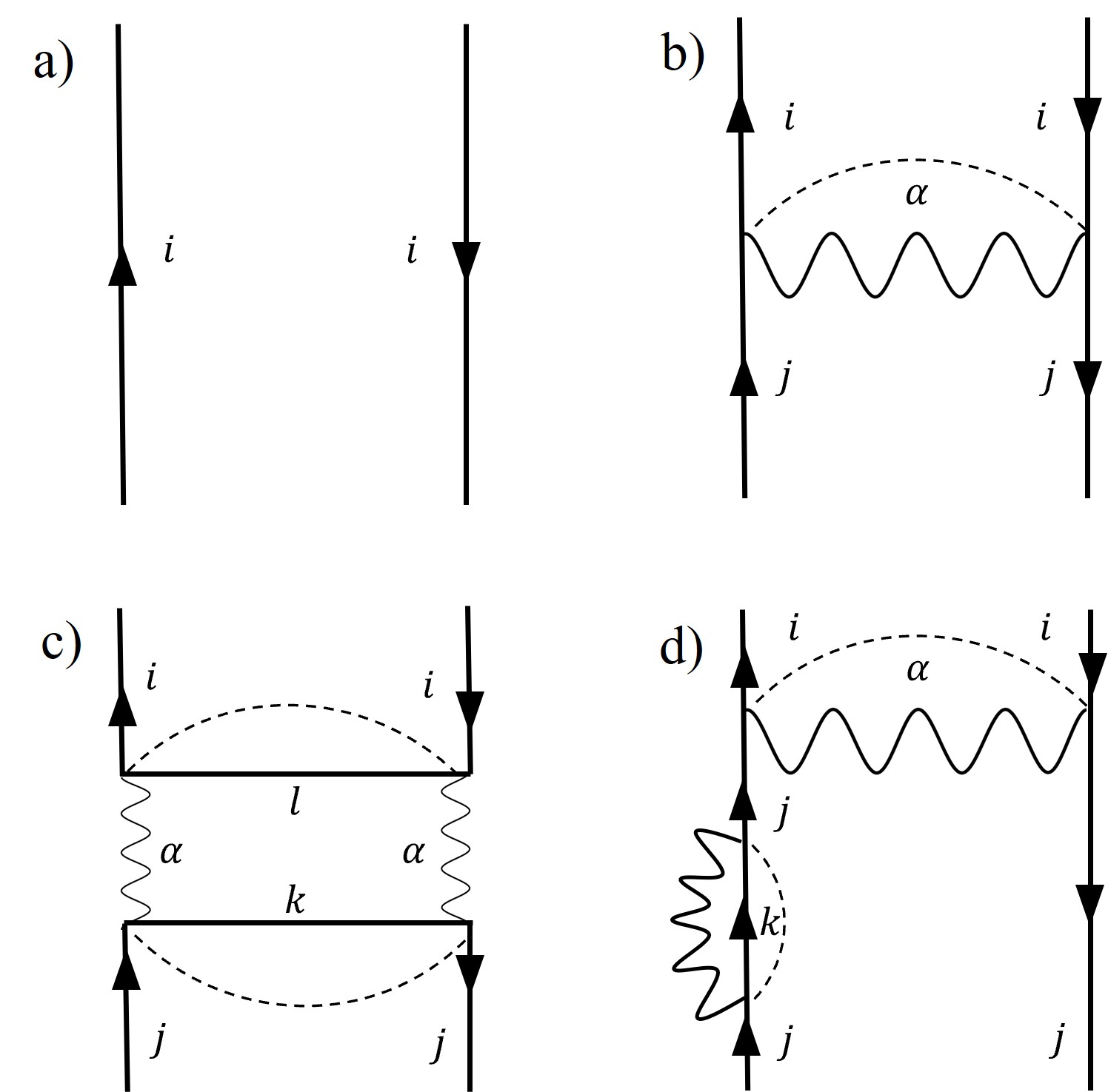}
\caption{Some diagrams at leading order in the large-$M$,$N$ expansion of the fermion OTOC. All propagators in these diagrams are bare, unlike in the figures in the main text.}
\label{appendixadiagrams}
\end{figure}

In this appendix we justify the claim that the leading order diagrams in the large-$M$,$N$ expansion of the OTOC~\eqref{eq:fermionOTOC} are those depicted in Fig.~\ref{fermionladder} by explicitly computing the factors of $M$ and $N$ up to second nonvanishing order in the interaction. At zeroth order, the fermion operators are free, so the first term is
\begin{align}
\frac{1}{M^2} &\sum_{ij}
\Tr [\rho^\frac{1}{2} \{c_i^0(t_1), c_j^{0\dagger}(0)\} \rho^\frac{1}{2} \{c_i^0(t_2), c_j^{0\dagger}(0)\}\da] \\
&= \frac{1}{M^2} \Tr [\rho^\frac{1}{2} G^{R,0}(t_1) \delta_{ij} \rho^\frac{1}{2} G^{A,0}(t_2) \delta_{ij}] \\
&= \frac{1}{M} G^{R,0}(t_1) G^{A,0}(t_2).
\end{align}
The superscript zeroes indicate free fermions and their free Green's functions. The summation over fermion flavors has produced one factor of $M$ for an overall factor of $1/M$ at this order. This term is represented by diagram (a) in Fig.~\ref{appendixadiagrams}. The fermion flavors on each rail are constrained to be the same, so the single factor of $M$ corresponds to the sum over the single fermion index $i$ in the diagram.

At first order in the interaction, all diagrams vanish because every term is proportional to the random coupling $t^\alpha_{ij}$ which has zero mean. All diagrams with an odd number of interaction vertices vanish similarly. This disorder averaging can be viewed as an additional field with a constant propagator which forces the two fermion flavors and the boson flavor at one end to match those at the other.

\begin{figure}[t]
\includegraphics[width=0.3\columnwidth]{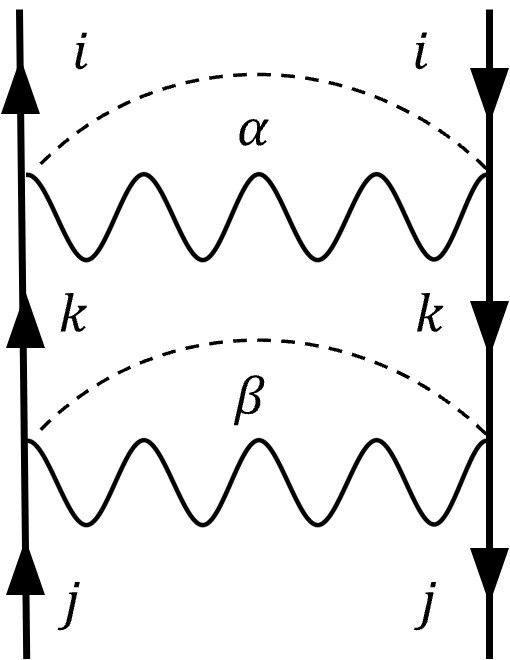}
\caption{A diagram with two Type-I rungs, included in the ladder series.}
\label{tworungs}
\end{figure}

At first nonvanishing order, the two vertices can either be placed on the same rail or with one on each rail. The former option, however, merely forms part of the self-energy for the fermion on that rail and is already included in Fig.~\ref{fermionladder} -- since those propagators are all dressed, the very first term in the sum contains this contribution. All the self-energy diagrams, by construction, are of leading order in $M$ and $N$, so these contributions to the OTOC are as well. The latter option is depicted in (b) of Fig.~\ref{appendixadiagrams} (the ``Type-I rung''). Compared to the zeroth order term, it contains one additional factor of $M$ from the new fermion sum, one additional factor of $N$ from the new boson sum, and a factor of $1/MN$ from the interaction. Hence this diagram is also leading in $M$ and $N$. In fact, the addition of each Type-I rung at any order has the same effect and does not change the $M$ and $N$ factors. This is true at all orders, so the diagram of Fig.~\ref{tworungs} for example is also dominant, and it is included in our ladder equation.

At second nonvanishing order, there are several ways of contracting the vertices, and the disorder average is over a product of four $t$'s, which can be decomposed by Isserlis's theorem into a sum of averages of all pairings of the $t$'s (diagrammatically, all the ways of pairing the vertices by dotted lines). Again many of these diagrams contain pieces which form part of the self-energy, such as (d) in Fig.~\ref{appendixadiagrams}, which is accounted for in Fig.~\ref{fermionladder} by the second term in the sum. Another possibility is two Type-I rungs in sequence. The important new diagram at this order is the ``Type-II'' rung (or ``box'') of (c) in Fig.~\ref{appendixadiagrams}. Compared to the zeroth order diagram, it has three more fermion sums (a factor of $M^3$), one more boson sum (a factor of $N$), and four more vertices (a factor of $1/M^2N^2$) for a combined factor of $M/N$. Hence this diagram is also of leading order, as are any diagrams obtained by appending more of these Type-II rungs.

\begin{figure}[h]
\includegraphics[width=0.9\columnwidth]{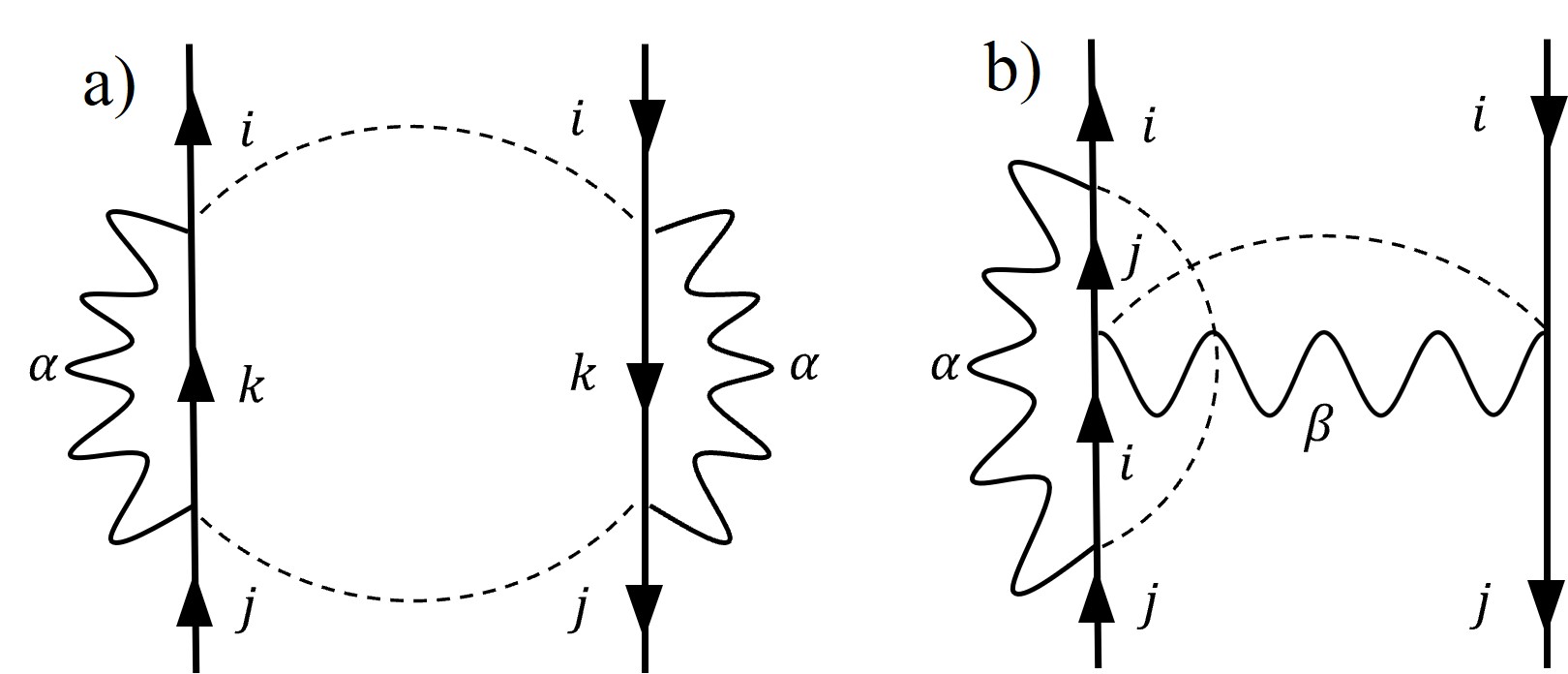}
\caption{Examples of subleading diagrams in the large-$M$,$N$ expansion of the fermion OTOC. All propagators are bare, unlike in the figures in the main text.}
\label{subleading}
\end{figure}

Notice that the pairing of vertices in the disorder average is important. For example, the Type-II rung must have the disorder average done as in (c). The other two possible pairings of the vertices force too many constraints on the flavors, and it can be easily verified that such terms are subleading in $M$ and/or $N$. Another example is shown in (a) of Fig.~\ref{subleading}: if the disorder average connected vertices on the same rung, this would be a relevant self-energy contribution, but with the inter-rung pairing as shown, it is suppressed by $1/N$.

There are a few other diagrams to check, including crossed rungs, other pairings in the disorder average, and diagram (b) of Fig.~\ref{subleading}, but it is not difficult to verify that they are all subleading. The dominant diagrams, at all orders, are precisely those achieved by iteratively tacking on Type-I and Type-II rungs to the zeroth order diagram as our ladder equation depicts.

\section{Feynman rule for the ladder diagrams in the OTOC}
\label{appendix:B}
\begin{figure}
\includegraphics[width=0.5\columnwidth]{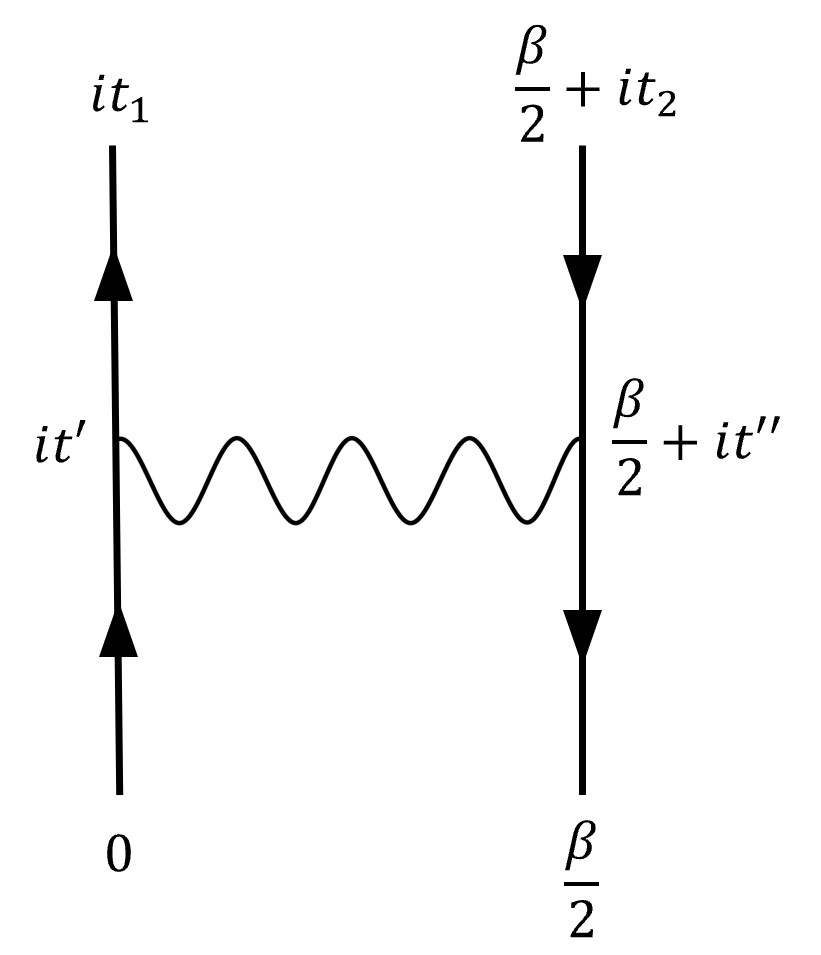}
\caption{The ladder diagram corresponding to~\eqref{eq:c1}.}
\label{onerungladder}
\end{figure}

In the interaction picture, the leading correction to the fermion OTOC $F_c(t_1,t_2)$ beyond self-energy effects is given by, using  Wick's theorem on the path integral along the complex time contour, 
\begin{widetext}
\begin{align}
F_1=&\frac{1}{M} \int_0^{t_1} \dd{t}'\int_0^{t_2} \dd{t}''
\wick{
\{[\c1 I\c2 (\c3 i t'),\c3 c^\dagger(i t_1)], \c1 c(0)\}\{[\c4 I\c2 (\c5 \beta/2+ i t''),\c5 c^\dagger(\beta/2+ it_2)],\c4 c(\beta/2)\}^\dagger 
}\nonumber\\
=& \frac{\omega_0^3}M \int_0^{t_1} \dd{t}'\int_0^{t_2} \dd{t}''
\wick{
\{[\c1 c^\dagger (i t') \c2 \phi(i t') \c3 c(i t'),\c3 c^\dagger(i t_1)],\c1 c(0) \}\{[\c4 c^\dagger (\beta/2+ it'') \c2 \phi(\beta/2+ it'') \c5 c( \beta/2+ i t''),\c5 c^\dagger(\beta/2+ it_2)],\c4 c(\beta/2)\}^\dagger,
}
\label{eq:c1}
 \end{align}
\end{widetext}
which corresponds to a ladder diagram with one rung, which we show in Fig.~\ref{onerungladder}. Here $I(t) \propto t c^\dagger(t)\phi(t)c(t)$ is the interaction vertex (with the flavor summations suppressed for brevity), which we used in the second step. The contraction is defined as the complex-time-contour-ordered correlator
\be
\wick {\c {\mathcal{O}_1}(t_1) \c {\mathcal{O}_2}(t_2)} = \langle T_{\rm cont}  \mathcal{O}_1(t_1) \mathcal{O}_2(t_2) \rangle,
\ee
where $T_{\rm cont}$ is the ordering operator on the complex time contour. Therefore for contractions across two real-time folds, we get Wightman propagators, e.g.,
\be
\label{eq:B3}
\wick{
\c2 \phi (0) \c2 \phi(\beta/2+it)= D^{lr}(t),
}
\ee

Importantly, as we mentioned in the main text, the (anti)-commutators in Eq.~\eqref{eq:c1} are not defined in the usual way, but via altering the order the operators on the complex time contour, i.e.,
\begin{align}
\label{eq:commu}
&\wick{\[ {\mathcal{O}_1}(it_1), {\mathcal{O}_2}(it_2)\]} _{\pm}\\
\equiv & \lim_{\epsilon\to 0} \mathcal{O}_1(\epsilon+it_1) \mathcal{O}_2(it_2) - \mathcal{O}_1(it_1) \mathcal{O}_2(\epsilon+it_2), \nonumber
\end{align}
where $[\cdots ]_+\equiv \{\cdots \}$.
Therefore in this notation
\be
\wick{
\{\c1 c(0),\c1 c^\dag(i t)\} \equiv  G^R(t),~~~\{\c1 c^\dagger(0),\c1 c(i t)\} \equiv  G^A(t).
}
\ee 
The nested (anti)-commutators in Eq.~\eqref{eq:c1} are defined using multiple $\epsilon_i\to 0$, making sure the limits are taken such that $\epsilon_i$ corresponding to inner (anti)-commutators tend to zero first.
It is easy to verify that the (anti)-commutator defined this way satisfies the same algebraic properties as the original version.

The expression in Eq.~\eqref{eq:c1} can be simplified by successively extracting contractions (c-numbers) from the nested (anti)-commutator. Importantly, when contracting (anti)-commutators, the only nonvanishing contribution comes from contractions that match the (anti)-commutation, i.e., between operators separated by a comma. In particular,
\be
\wick{
 [ {\bar a},\c3 b]_{\pm}\c3{\bar b} =0,
}
\label{eq:c2}
\ee
because the order between $b$ and $\bar b$ is not altered by the $\epsilon$-prescription above, and the two terms forming the commutator defined in Eq.~\eqref{eq:commu} cancel.
As an example for nested commutators,
\be
\wick{
\{\c1 a,[\c1{\bar a} \c2{\bar b}, \c2 b]\} = - \{\c4 a, \{\c4 {\bar a},\c3 b\}\c3{\bar b}\} +\{\c4 a, \c4 {\bar a}\{\c3 b, \c3{\bar b}\} \} = \{\c4 a, \c4 {\bar a}\}\{\c3 b, \c3{\bar b}\},
}
\ee
where in the first step we used the mathematical identity $[\bar a \bar b, b] \equiv -\{\bar a, b\}\bar b + \bar a \{\bar b,\bar b\}$, and the first term in the middle expression vanishes because of \eqref{eq:c2}. Combining this recipe with Eq.~\eqref{eq:B3}, it can be shown in general that for an OTOC with nested commutators, one can successively replace contractions separated by a comma with the retarded or advanced Green's functions, and replace contractions across real-time folds via Wightman propagators. This is the analog of Feynman rules for OTOCs.

Applying the Feynman rules, we get
\begin{align}
F_1=&\frac{\omega_0^3}{M}\int_{-\infty}^{\infty} \dd{t}'\int_{-\infty}^{\infty} \dd{t}''  G^R(t') G^R(t_1-t') D^{lr}(t''-t')\nonumber \\
&\times G^A(t'')G^A(t_2-t'').
\end{align}

Higher-order ladder diagrams are obtained by replacing $\int_0^{t_1} \dd{t}' [I(it'),c^\dagger(it_1)]$ with 
\be
\label{eq:higherorderladder}
 \int_0^{t_1} \dd{t}_n'\cdots \int_{0}^{\dd{t}_2'} \dd{t}_1' [I(it_n'),[\cdots, [I(it_1'),c^\dagger(it_1)],
\ee
which can then be contracted using the same procedure.
It is straightforward to see that the Feynman rule
for the ladder diagrams in the OTOC is to assign a retarded (advanced) Green's function to all vertical lines, and a Wightman correlator to the horizontal lines.

\section{Leading order diagrams involving $F_{\bar c}$ from Feynman rules}
\label{appendix:C}

At low orders in the expansion of the OTOC, one finds retarded Green's functions on one rail of the ladders and advanced ones on the other. However, it is possible for both retarded and advanced Green's functions to appear on either rail. This first becomes relevant to the ``anomalous" OTOC $F_{\bar c}$ at second order in perturbation theory, and relevant to the OTOC $F_c$ at fourth order in perturbation theory.

\begin{figure}[t]
\includegraphics[width=0.7\columnwidth]{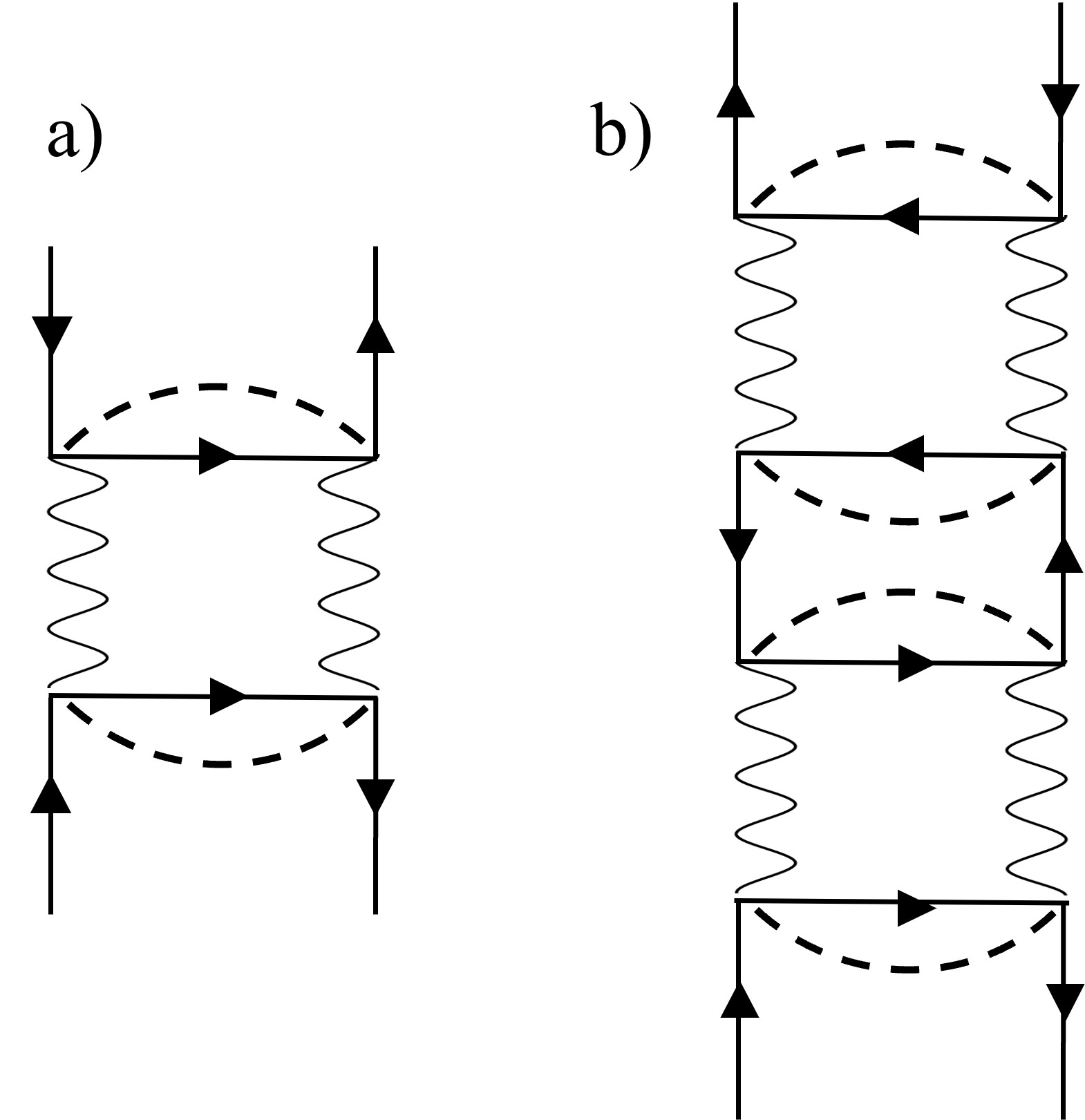}
\caption{(a) A contribution to the ``anomalous'' OTOC at second order. (b) A contribution to the original OTOC at fourth order, which contains (a) as a subdiagram. All propagators are bare.}
\label{appendixcdiagrams}
\end{figure}

We begin with
\be
F_{\bar c} \sim \Tr[\rho^{1/2}\{\tilde{c}_i(it_1), \tilde{c}_j(0)\} \rho^{1/2}\{\tilde{c}_i(it_2), \tilde{c}_j(0)\}\da]
\ee
which is depicted on the left side of the second equation in Fig.~\ref{fermionladder}. The tildes indicate that these are Heisenberg picture operators, which are related to the interaction picture operators $c(\tau)$ by $\tilde{c}(\tau) = U\da c(\tau) U$, where
\be
U = \mathcal{T} \exp \qty( -\frac{i}{\sqrt{MN}} t_{k k'}^\alpha \int_0^\tau \dd{\tau'} c_k\da(\tau') c_{k'}(\tau') \phi_\alpha(\tau'))
\ee
with flavor summations implied. If we expand $U\da c(\tau) U$ to second order in the random coupling, we find
\begin{widetext}
\be
[U\da c_i(t_1) U]^{(2)} = \frac{-1}{MN} t_{k k'}^\alpha t_{l l'}^\beta \int_0^t \dd{t'} \int_0^{t'} \dd{t''}
[c\da_l (it'') c_{l'}(it'') \phi_\beta(it''), [c\da_k (it') c_{k'}(it') \phi_\alpha(it'), c_i(it_1)] ]
\ee
as detailed in~\cite{PhysRevD.96.065005} and~\cite{PhysRevLett.129.060601}, and as claimed in Eq.~\eqref{eq:higherorderladder}. Suppressing prefactors, integrations, and flavors indices for brevity, there is a diagram (Fig.~\ref{appendixcdiagrams}(a)) corresponding to the contraction
\bigskip
\begin{align}
&F_{\bar c}=\frac{\omega_0^6}{M}\times \nonumber\\
&\wick{
\int \!\! \dd{t'}\!\dd{t''}\!\dd{t'''}\!\dd{t''''} \{ [ \c2 c\da \c8 c \c5 \phi(it''),
[ \c1 c\da \c7 c \c5 \phi(t'), \c1 c_i(it_1)]], \c2 c_j(0) \}
\{ [ \c8 c\da \c4 c \c6 \phi({\beta}/{2}+it''''),
[ \c7 c\da \c3 c \c6 \phi({\beta}/{2}+it'''), \c3 c_i\da({\beta}/{2}+it_2)]], \c4 c_j\da({\beta}/{2}) \} ]
}\nonumber\\
&+\cdots,
\label{longcontraction}
\end{align}
where we reiterate that the notation with both commutators and contractions is explained in Appendix \ref{appendix:B}. Applying the Feynman rules developed in Appendix \ref{appendix:B}, we find that
\eqref{longcontraction} produces an integral with the structure
\begin{align}
F_{\bar c} = \frac{\omega_0^6}{M}
&\int \!\! \dd{t'}\!\dd{t''}\!\dd{t'''}\!\dd{t''''}
G^R(t_1-t') D^R(t'-t'') G^A(t'') G^r(t'-t''') G^r(t''-t'''') G^A(t_2-t''') D^R(t'''-t'''') G^R(t''''). \nonumber\\
&+ \cdots
\end{align}
\end{widetext}
This diagram is the last term of the second equation in Fig.~\ref{fermionladder} with the shaded rectangle removed. As a ladder diagram similar to that in Fig.~\ref{appendixadiagrams}(c), it is straightforward to show that this diagram is of $O(1/M)$, i.e., at the same order as other leading diagrams in the series. This contribution is included in the integral equation \eqref{ladder}, as can be seen via an iterative expansion.

As is indicated in the top equation of Fig.~\ref{fermionladder}, the existence of $F_{\bar c}$ leads to an additional contribution to $F_c$, which at leading order is diagrammatically shown in Fig.~\ref{appendixcdiagrams}(b). Again this diagram is at $O(1/M)$ and can be explicitly written in contractions just like  Eq.~\eqref{longcontraction}. Writing it explicitly in terms of contracted nested commutators will unfortunately take too much space, but the corresponding Feynman diagram is obtained by combining two leading diagrams for $F_{\bar c}$. Applying the Feynman rule developed in Appendix \ref{appendix:B}, we get  
\begin{widetext}
\begin{align}
F_c  =& \cdots + \frac{\omega_0^{12}}{M} \int \!\! \dd{\bar t'}\!\dd{\bar t''}\!\dd{\bar t'''}\!\dd{\bar t''''}
G^R(t_1-\bar t') D^R(\bar t'-\bar t'')G^r(\bar t'-\bar t''') G^r(\bar t''-\bar t'''') G^A(t_2-\bar t''') D^R(\bar t'''-\bar t'''') \nonumber\\
&\int \!\! \dd{t'}\!\dd{t''}\!\dd{t'''}\!\dd{t''''}
G^R(\bar t''''-t') D^R(t'-t'') G^A(t'') G^l(t'-t''') G^l(t''-t'''') G^A(\bar t''-t''') D^R(t'''-t'''') G^R(t'''') \nonumber\\
&+ \cdots,
\end{align}
which is also obtained by iteratively expanding the integral equation \eqref{ladder}.
\end{widetext}

\bibliography{ReferencesYSYK2022}

\begin{thebibliography}{41}%
\makeatletter
\providecommand \@ifxundefined [1]{%
 \@ifx{#1\undefined}
}%
\providecommand \@ifnum [1]{%
 \ifnum #1\expandafter \@firstoftwo
 \else \expandafter \@secondoftwo
 \fi
}%
\providecommand \@ifx [1]{%
 \ifx #1\expandafter \@firstoftwo
 \else \expandafter \@secondoftwo
 \fi
}%
\providecommand \natexlab [1]{#1}%
\providecommand \enquote  [1]{``#1''}%
\providecommand \bibnamefont  [1]{#1}%
\providecommand \bibfnamefont [1]{#1}%
\providecommand \citenamefont [1]{#1}%
\providecommand \href@noop [0]{\@secondoftwo}%
\providecommand \href [0]{\begingroup \@sanitize@url \@href}%
\providecommand \@href[1]{\@@startlink{#1}\@@href}%
\providecommand \@@href[1]{\endgroup#1\@@endlink}%
\providecommand \@sanitize@url [0]{\catcode `\\12\catcode `\$12\catcode
  `\&12\catcode `\#12\catcode `\^12\catcode `\_12\catcode `\%12\relax}%
\providecommand \@@startlink[1]{}%
\providecommand \@@endlink[0]{}%
\providecommand \url  [0]{\begingroup\@sanitize@url \@url }%
\providecommand \@url [1]{\endgroup\@href {#1}{\urlprefix }}%
\providecommand \urlprefix  [0]{URL }%
\providecommand \Eprint [0]{\href }%
\providecommand \doibase [0]{https://doi.org/}%
\providecommand \selectlanguage [0]{\@gobble}%
\providecommand \bibinfo  [0]{\@secondoftwo}%
\providecommand \bibfield  [0]{\@secondoftwo}%
\providecommand \translation [1]{[#1]}%
\providecommand \BibitemOpen [0]{}%
\providecommand \bibitemStop [0]{}%
\providecommand \bibitemNoStop [0]{.\EOS\space}%
\providecommand \EOS [0]{\spacefactor3000\relax}%
\providecommand \BibitemShut  [1]{\csname bibitem#1\endcsname}%
\let\auto@bib@innerbib\@empty
\bibitem [{\citenamefont {Kitaev}()}]{kitaevsyk}%
  \BibitemOpen
  \bibfield  {author} {\bibinfo {author} {\bibfnamefont {A.}~\bibnamefont
  {Kitaev}},\ }\href@noop {} {\bibinfo {title} {A simple model of quantum
  holography}},\ \bibinfo {howpublished} {KITP strings seminar and Entanglement
  2015 program (Feb 12, Apr 7, and May 27, 2015)},\ \bibinfo {note}
  {https://online.kitp.ucsb.edu/online/entangled15/}\BibitemShut {NoStop}%
\bibitem [{\citenamefont {Chowdhury}\ \emph {et~al.}(2022)\citenamefont
  {Chowdhury}, \citenamefont {Georges}, \citenamefont {Parcollet},\ and\
  \citenamefont {Sachdev}}]{Chowdhury:2021qpy}%
  \BibitemOpen
  \bibfield  {author} {\bibinfo {author} {\bibfnamefont {D.}~\bibnamefont
  {Chowdhury}}, \bibinfo {author} {\bibfnamefont {A.}~\bibnamefont {Georges}},
  \bibinfo {author} {\bibfnamefont {O.}~\bibnamefont {Parcollet}},\ and\
  \bibinfo {author} {\bibfnamefont {S.}~\bibnamefont {Sachdev}},\ }\bibfield
  {title} {\bibinfo {title} {{Sachdev-Ye-Kitaev models and beyond: Window into
  non-Fermi liquids}},\ }\href {https://doi.org/10.1103/RevModPhys.94.035004}
  {\bibfield  {journal} {\bibinfo  {journal} {Rev. Mod. Phys.}\ }\textbf
  {\bibinfo {volume} {94}},\ \bibinfo {pages} {035004} (\bibinfo {year}
  {2022})},\ \Eprint {https://arxiv.org/abs/2109.05037} {arXiv:2109.05037
  [cond-mat.str-el]} \BibitemShut {NoStop}%
\bibitem [{\citenamefont {Fu}\ \emph {et~al.}(2017)\citenamefont {Fu},
  \citenamefont {Gaiotto}, \citenamefont {Maldacena},\ and\ \citenamefont
  {Sachdev}}]{PhysRevD.95.026009}%
  \BibitemOpen
  \bibfield  {author} {\bibinfo {author} {\bibfnamefont {W.}~\bibnamefont
  {Fu}}, \bibinfo {author} {\bibfnamefont {D.}~\bibnamefont {Gaiotto}},
  \bibinfo {author} {\bibfnamefont {J.}~\bibnamefont {Maldacena}},\ and\
  \bibinfo {author} {\bibfnamefont {S.}~\bibnamefont {Sachdev}},\ }\bibfield
  {title} {\bibinfo {title} {Supersymmetric sachdev-ye-kitaev models},\ }\href
  {https://doi.org/10.1103/PhysRevD.95.026009} {\bibfield  {journal} {\bibinfo
  {journal} {Phys. Rev. D}\ }\textbf {\bibinfo {volume} {95}},\ \bibinfo
  {pages} {026009} (\bibinfo {year} {2017})}\BibitemShut {NoStop}%
\bibitem [{\citenamefont {Rosenhaus}(2019)}]{Rosenhaus:2018dtp}%
  \BibitemOpen
  \bibfield  {author} {\bibinfo {author} {\bibfnamefont {V.}~\bibnamefont
  {Rosenhaus}},\ }\bibfield  {title} {\bibinfo {title} {{An introduction to the
  SYK model}},\ }\href {https://doi.org/10.1088/1751-8121/ab2ce1} {\bibfield
  {journal} {\bibinfo  {journal} {J. Phys. A}\ }\textbf {\bibinfo {volume}
  {52}},\ \bibinfo {pages} {323001} (\bibinfo {year} {2019})},\ \Eprint
  {https://arxiv.org/abs/1807.03334} {arXiv:1807.03334 [hep-th]} \BibitemShut
  {NoStop}%
\bibitem [{\citenamefont {Maldacena}\ and\ \citenamefont
  {Stanford}(2016)}]{PhysRevD.94.106002}%
  \BibitemOpen
  \bibfield  {author} {\bibinfo {author} {\bibfnamefont {J.}~\bibnamefont
  {Maldacena}}\ and\ \bibinfo {author} {\bibfnamefont {D.}~\bibnamefont
  {Stanford}},\ }\bibfield  {title} {\bibinfo {title} {Remarks on the
  sachdev-ye-kitaev model},\ }\href
  {https://doi.org/10.1103/PhysRevD.94.106002} {\bibfield  {journal} {\bibinfo
  {journal} {Phys. Rev. D}\ }\textbf {\bibinfo {volume} {94}},\ \bibinfo
  {pages} {106002} (\bibinfo {year} {2016})}\BibitemShut {NoStop}%
\bibitem [{\citenamefont {Kitaev}\ and\ \citenamefont
  {Suh}(2018)}]{Kitaev:2017awl}%
  \BibitemOpen
  \bibfield  {author} {\bibinfo {author} {\bibfnamefont {A.}~\bibnamefont
  {Kitaev}}\ and\ \bibinfo {author} {\bibfnamefont {S.~J.}\ \bibnamefont
  {Suh}},\ }\bibfield  {title} {\bibinfo {title} {{The soft mode in the
  Sachdev-Ye-Kitaev model and its gravity dual}},\ }\href
  {https://doi.org/10.1007/JHEP05(2018)183} {\bibfield  {journal} {\bibinfo
  {journal} {JHEP}\ }\textbf {\bibinfo {volume} {05}},\ \bibinfo {pages}
  {183}},\ \Eprint {https://arxiv.org/abs/1711.08467} {arXiv:1711.08467
  [hep-th]} \BibitemShut {NoStop}%
\bibitem [{\citenamefont {{Patel}}\ \emph {et~al.}(2022)\citenamefont
  {{Patel}}, \citenamefont {{Guo}}, \citenamefont {{Esterlis}},\ and\
  \citenamefont {{Sachdev}}}]{2022arXiv220304990P}%
  \BibitemOpen
  \bibfield  {author} {\bibinfo {author} {\bibfnamefont {A.~A.}\ \bibnamefont
  {{Patel}}}, \bibinfo {author} {\bibfnamefont {H.}~\bibnamefont {{Guo}}},
  \bibinfo {author} {\bibfnamefont {I.}~\bibnamefont {{Esterlis}}},\ and\
  \bibinfo {author} {\bibfnamefont {S.}~\bibnamefont {{Sachdev}}},\ }\bibfield
  {title} {\bibinfo {title} {{Universal theory of strange metals from spatially
  random interactions}},\ }\href@noop {} {\bibfield  {journal} {\bibinfo
  {journal} {arXiv e-prints}\ ,\ \bibinfo {eid} {arXiv:2203.04990}} (\bibinfo
  {year} {2022})},\ \Eprint {https://arxiv.org/abs/2203.04990}
  {arXiv:2203.04990 [cond-mat.str-el]} \BibitemShut {NoStop}%
\bibitem [{\citenamefont {Esterlis}\ \emph {et~al.}(2021)\citenamefont
  {Esterlis}, \citenamefont {Guo}, \citenamefont {Patel},\ and\ \citenamefont
  {Sachdev}}]{PhysRevB.103.235129}%
  \BibitemOpen
  \bibfield  {author} {\bibinfo {author} {\bibfnamefont {I.}~\bibnamefont
  {Esterlis}}, \bibinfo {author} {\bibfnamefont {H.}~\bibnamefont {Guo}},
  \bibinfo {author} {\bibfnamefont {A.~A.}\ \bibnamefont {Patel}},\ and\
  \bibinfo {author} {\bibfnamefont {S.}~\bibnamefont {Sachdev}},\ }\bibfield
  {title} {\bibinfo {title} {Large-$n$ theory of critical fermi surfaces},\
  }\href {https://doi.org/10.1103/PhysRevB.103.235129} {\bibfield  {journal}
  {\bibinfo  {journal} {Phys. Rev. B}\ }\textbf {\bibinfo {volume} {103}},\
  \bibinfo {pages} {235129} (\bibinfo {year} {2021})}\BibitemShut {NoStop}%
\bibitem [{\citenamefont {Aldape}\ \emph {et~al.}(2022)\citenamefont {Aldape},
  \citenamefont {Cookmeyer}, \citenamefont {Patel},\ and\ \citenamefont
  {Altman}}]{PhysRevB.105.235111}%
  \BibitemOpen
  \bibfield  {author} {\bibinfo {author} {\bibfnamefont {E.~E.}\ \bibnamefont
  {Aldape}}, \bibinfo {author} {\bibfnamefont {T.}~\bibnamefont {Cookmeyer}},
  \bibinfo {author} {\bibfnamefont {A.~A.}\ \bibnamefont {Patel}},\ and\
  \bibinfo {author} {\bibfnamefont {E.}~\bibnamefont {Altman}},\ }\bibfield
  {title} {\bibinfo {title} {Solvable theory of a strange metal at the
  breakdown of a heavy fermi liquid},\ }\href
  {https://doi.org/10.1103/PhysRevB.105.235111} {\bibfield  {journal} {\bibinfo
   {journal} {Phys. Rev. B}\ }\textbf {\bibinfo {volume} {105}},\ \bibinfo
  {pages} {235111} (\bibinfo {year} {2022})}\BibitemShut {NoStop}%
\bibitem [{\citenamefont {Hussey}\ \emph {et~al.}(1998)\citenamefont {Hussey},
  \citenamefont {Mackenzie}, \citenamefont {Cooper}, \citenamefont {Maeno},
  \citenamefont {Nishizaki},\ and\ \citenamefont {Fujita}}]{PhysRevB.57.5505}%
  \BibitemOpen
  \bibfield  {author} {\bibinfo {author} {\bibfnamefont {N.~E.}\ \bibnamefont
  {Hussey}}, \bibinfo {author} {\bibfnamefont {A.~P.}\ \bibnamefont
  {Mackenzie}}, \bibinfo {author} {\bibfnamefont {J.~R.}\ \bibnamefont
  {Cooper}}, \bibinfo {author} {\bibfnamefont {Y.}~\bibnamefont {Maeno}},
  \bibinfo {author} {\bibfnamefont {S.}~\bibnamefont {Nishizaki}},\ and\
  \bibinfo {author} {\bibfnamefont {T.}~\bibnamefont {Fujita}},\ }\bibfield
  {title} {\bibinfo {title} {Normal-state magnetoresistance of
  ${\mathrm{sr}}_{2}{\mathrm{ruo}}_{4}$},\ }\href
  {https://doi.org/10.1103/PhysRevB.57.5505} {\bibfield  {journal} {\bibinfo
  {journal} {Phys. Rev. B}\ }\textbf {\bibinfo {volume} {57}},\ \bibinfo
  {pages} {5505} (\bibinfo {year} {1998})}\BibitemShut {NoStop}%
\bibitem [{\citenamefont {Stewart}(2001)}]{RevModPhys.73.797}%
  \BibitemOpen
  \bibfield  {author} {\bibinfo {author} {\bibfnamefont {G.~R.}\ \bibnamefont
  {Stewart}},\ }\bibfield  {title} {\bibinfo {title} {Non-fermi-liquid behavior
  in $d$- and $f$-electron metals},\ }\href
  {https://doi.org/10.1103/RevModPhys.73.797} {\bibfield  {journal} {\bibinfo
  {journal} {Rev. Mod. Phys.}\ }\textbf {\bibinfo {volume} {73}},\ \bibinfo
  {pages} {797} (\bibinfo {year} {2001})}\BibitemShut {NoStop}%
\bibitem [{\citenamefont {{Keimer}}\ \emph {et~al.}(2014)\citenamefont
  {{Keimer}}, \citenamefont {{Kivelson}}, \citenamefont {{Norman}},
  \citenamefont {{Uchida}},\ and\ \citenamefont
  {{Zaanen}}}]{2014arXiv1409.4673K}%
  \BibitemOpen
  \bibfield  {author} {\bibinfo {author} {\bibfnamefont {B.}~\bibnamefont
  {{Keimer}}}, \bibinfo {author} {\bibfnamefont {S.~A.}\ \bibnamefont
  {{Kivelson}}}, \bibinfo {author} {\bibfnamefont {M.~R.}\ \bibnamefont
  {{Norman}}}, \bibinfo {author} {\bibfnamefont {S.}~\bibnamefont {{Uchida}}},\
  and\ \bibinfo {author} {\bibfnamefont {J.}~\bibnamefont {{Zaanen}}},\
  }\bibfield  {title} {\bibinfo {title} {{High Temperature Superconductivity in
  the Cuprates}},\ }\href@noop {} {\bibfield  {journal} {\bibinfo  {journal}
  {arXiv e-prints}\ ,\ \bibinfo {eid} {arXiv:1409.4673}} (\bibinfo {year}
  {2014})},\ \Eprint {https://arxiv.org/abs/1409.4673} {arXiv:1409.4673
  [cond-mat.supr-con]} \BibitemShut {NoStop}%
\bibitem [{\citenamefont {Varma}\ \emph {et~al.}(1989)\citenamefont {Varma},
  \citenamefont {Littlewood}, \citenamefont {Schmitt-Rink}, \citenamefont
  {Abrahams},\ and\ \citenamefont {Ruckenstein}}]{PhysRevLett.63.1996}%
  \BibitemOpen
  \bibfield  {author} {\bibinfo {author} {\bibfnamefont {C.~M.}\ \bibnamefont
  {Varma}}, \bibinfo {author} {\bibfnamefont {P.~B.}\ \bibnamefont
  {Littlewood}}, \bibinfo {author} {\bibfnamefont {S.}~\bibnamefont
  {Schmitt-Rink}}, \bibinfo {author} {\bibfnamefont {E.}~\bibnamefont
  {Abrahams}},\ and\ \bibinfo {author} {\bibfnamefont {A.~E.}\ \bibnamefont
  {Ruckenstein}},\ }\bibfield  {title} {\bibinfo {title} {Phenomenology of the
  normal state of cu-o high-temperature superconductors},\ }\href
  {https://doi.org/10.1103/PhysRevLett.63.1996} {\bibfield  {journal} {\bibinfo
   {journal} {Phys. Rev. Lett.}\ }\textbf {\bibinfo {volume} {63}},\ \bibinfo
  {pages} {1996} (\bibinfo {year} {1989})}\BibitemShut {NoStop}%
\bibitem [{\citenamefont {Cha}\ \emph {et~al.}(2020)\citenamefont {Cha},
  \citenamefont {Wentzell}, \citenamefont {Parcollet}, \citenamefont
  {Georges},\ and\ \citenamefont {Kim}}]{Cha18341}%
  \BibitemOpen
  \bibfield  {author} {\bibinfo {author} {\bibfnamefont {P.}~\bibnamefont
  {Cha}}, \bibinfo {author} {\bibfnamefont {N.}~\bibnamefont {Wentzell}},
  \bibinfo {author} {\bibfnamefont {O.}~\bibnamefont {Parcollet}}, \bibinfo
  {author} {\bibfnamefont {A.}~\bibnamefont {Georges}},\ and\ \bibinfo {author}
  {\bibfnamefont {E.-A.}\ \bibnamefont {Kim}},\ }\bibfield  {title} {\bibinfo
  {title} {Linear resistivity and sachdev-ye-kitaev (syk) spin liquid behavior
  in a quantum critical metal with spin-1/2 fermions},\ }\href
  {https://doi.org/10.1073/pnas.2003179117} {\bibfield  {journal} {\bibinfo
  {journal} {Proceedings of the National Academy of Sciences}\ }\textbf
  {\bibinfo {volume} {117}},\ \bibinfo {pages} {18341} (\bibinfo {year}
  {2020})},\ \Eprint
  {https://arxiv.org/abs/https://www.pnas.org/content/117/31/18341.full.pdf}
  {https://www.pnas.org/content/117/31/18341.full.pdf} \BibitemShut {NoStop}%
\bibitem [{\citenamefont {Guo}\ \emph {et~al.}(2020)\citenamefont {Guo},
  \citenamefont {Gu},\ and\ \citenamefont {Sachdev}}]{Guo:2020aog}%
  \BibitemOpen
  \bibfield  {author} {\bibinfo {author} {\bibfnamefont {H.}~\bibnamefont
  {Guo}}, \bibinfo {author} {\bibfnamefont {Y.}~\bibnamefont {Gu}},\ and\
  \bibinfo {author} {\bibfnamefont {S.}~\bibnamefont {Sachdev}},\ }\bibfield
  {title} {\bibinfo {title} {{Linear in temperature resistivity in the limit of
  zero temperature from the time reparameterization soft mode}},\ }\href
  {https://doi.org/10.1016/j.aop.2020.168202} {\bibfield  {journal} {\bibinfo
  {journal} {Annals Phys.}\ }\textbf {\bibinfo {volume} {418}},\ \bibinfo
  {pages} {168202} (\bibinfo {year} {2020})},\ \Eprint
  {https://arxiv.org/abs/2004.05182} {arXiv:2004.05182 [cond-mat.str-el]}
  \BibitemShut {NoStop}%
\bibitem [{\citenamefont
  {Lee}(2018)}]{doi:10.1146/annurev-conmatphys-031016-025531}%
  \BibitemOpen
  \bibfield  {author} {\bibinfo {author} {\bibfnamefont {S.-S.}\ \bibnamefont
  {Lee}},\ }\bibfield  {title} {\bibinfo {title} {Recent developments in
  non-fermi liquid theory},\ }\href
  {https://doi.org/10.1146/annurev-conmatphys-031016-025531} {\bibfield
  {journal} {\bibinfo  {journal} {Annual Review of Condensed Matter Physics}\
  }\textbf {\bibinfo {volume} {9}},\ \bibinfo {pages} {227} (\bibinfo {year}
  {2018})},\ \Eprint
  {https://arxiv.org/abs/https://doi.org/10.1146/annurev-conmatphys-031016-025531}
  {https://doi.org/10.1146/annurev-conmatphys-031016-025531} \BibitemShut
  {NoStop}%
\bibitem [{\citenamefont {Wang}(2020)}]{Wang2020SolvableSQ}%
  \BibitemOpen
  \bibfield  {author} {\bibinfo {author} {\bibfnamefont {Y.}~\bibnamefont
  {Wang}},\ }\bibfield  {title} {\bibinfo {title} {Solvable strong-coupling
  quantum-dot model with a non-fermi-liquid pairing transition.},\ }\href@noop
  {} {\bibfield  {journal} {\bibinfo  {journal} {Physical review letters}\
  }\textbf {\bibinfo {volume} {124 1}},\ \bibinfo {pages} {017002} (\bibinfo
  {year} {2020})}\BibitemShut {NoStop}%
\bibitem [{\citenamefont {Esterlis}\ and\ \citenamefont
  {Schmalian}(2019)}]{PhysRevB.100.115132}%
  \BibitemOpen
  \bibfield  {author} {\bibinfo {author} {\bibfnamefont {I.}~\bibnamefont
  {Esterlis}}\ and\ \bibinfo {author} {\bibfnamefont {J.}~\bibnamefont
  {Schmalian}},\ }\bibfield  {title} {\bibinfo {title} {Cooper pairing of
  incoherent electrons: An electron-phonon version of the sachdev-ye-kitaev
  model},\ }\href {https://doi.org/10.1103/PhysRevB.100.115132} {\bibfield
  {journal} {\bibinfo  {journal} {Phys. Rev. B}\ }\textbf {\bibinfo {volume}
  {100}},\ \bibinfo {pages} {115132} (\bibinfo {year} {2019})}\BibitemShut
  {NoStop}%
\bibitem [{\citenamefont {{Choi}}\ \emph {et~al.}(2022)\citenamefont {{Choi}},
  \citenamefont {{Tavakol}},\ and\ \citenamefont
  {{Kim}}}]{2022ScPP...12..151C}%
  \BibitemOpen
  \bibfield  {author} {\bibinfo {author} {\bibfnamefont {W.}~\bibnamefont
  {{Choi}}}, \bibinfo {author} {\bibfnamefont {O.}~\bibnamefont {{Tavakol}}},\
  and\ \bibinfo {author} {\bibfnamefont {Y.~B.}\ \bibnamefont {{Kim}}},\
  }\bibfield  {title} {\bibinfo {title} {{Pairing instabilities of the
  Yukawa-SYK models with controlled fermion incoherence}},\ }\href
  {https://doi.org/10.21468/SciPostPhys.12.5.151} {\bibfield  {journal}
  {\bibinfo  {journal} {SciPost Physics}\ }\textbf {\bibinfo {volume} {12}},\
  \bibinfo {eid} {151} (\bibinfo {year} {2022})},\ \Eprint
  {https://arxiv.org/abs/2110.02968} {arXiv:2110.02968 [cond-mat.str-el]}
  \BibitemShut {NoStop}%
\bibitem [{\citenamefont {Hauck}\ \emph {et~al.}(2020)\citenamefont {Hauck},
  \citenamefont {Klug}, \citenamefont {Esterlis},\ and\ \citenamefont
  {Schmalian}}]{HAUCK2020168120}%
  \BibitemOpen
  \bibfield  {author} {\bibinfo {author} {\bibfnamefont {D.}~\bibnamefont
  {Hauck}}, \bibinfo {author} {\bibfnamefont {M.~J.}\ \bibnamefont {Klug}},
  \bibinfo {author} {\bibfnamefont {I.}~\bibnamefont {Esterlis}},\ and\
  \bibinfo {author} {\bibfnamefont {J.}~\bibnamefont {Schmalian}},\ }\bibfield
  {title} {\bibinfo {title} {Eliashberg equations for an electron–phonon
  version of the sachdev–ye–kitaev model: Pair breaking in non-fermi liquid
  superconductors},\ }\href
  {https://doi.org/https://doi.org/10.1016/j.aop.2020.168120} {\bibfield
  {journal} {\bibinfo  {journal} {Annals of Physics}\ }\textbf {\bibinfo
  {volume} {417}},\ \bibinfo {pages} {168120} (\bibinfo {year} {2020})},\
  \bibinfo {note} {eliashberg theory at 60: Strong-coupling superconductivity
  and beyond}\BibitemShut {NoStop}%
\bibitem [{\citenamefont {Kim}\ \emph {et~al.}(2021)\citenamefont {Kim},
  \citenamefont {Altman},\ and\ \citenamefont {Cao}}]{PhysRevB.103.L081113}%
  \BibitemOpen
  \bibfield  {author} {\bibinfo {author} {\bibfnamefont {J.}~\bibnamefont
  {Kim}}, \bibinfo {author} {\bibfnamefont {E.}~\bibnamefont {Altman}},\ and\
  \bibinfo {author} {\bibfnamefont {X.}~\bibnamefont {Cao}},\ }\bibfield
  {title} {\bibinfo {title} {Dirac fast scramblers},\ }\href
  {https://doi.org/10.1103/PhysRevB.103.L081113} {\bibfield  {journal}
  {\bibinfo  {journal} {Phys. Rev. B}\ }\textbf {\bibinfo {volume} {103}},\
  \bibinfo {pages} {L081113} (\bibinfo {year} {2021})}\BibitemShut {NoStop}%
\bibitem [{\citenamefont {Schmalian}(2022)}]{SchmalianHolographic2022}%
  \BibitemOpen
  \bibfield  {author} {\bibinfo {author} {\bibfnamefont {J.}~\bibnamefont
  {Schmalian}},\ }\href {https://doi.org/10.48550/ARXIV.2209.00474} {\bibinfo
  {title} {Holographic superconductivity of a critical fermi surface}}
  (\bibinfo {year} {2022})\BibitemShut {NoStop}%
\bibitem [{\citenamefont {Wang}\ and\ \citenamefont
  {Chubukov}(2020)}]{Wang:2020dtj}%
  \BibitemOpen
  \bibfield  {author} {\bibinfo {author} {\bibfnamefont {Y.}~\bibnamefont
  {Wang}}\ and\ \bibinfo {author} {\bibfnamefont {A.~V.}\ \bibnamefont
  {Chubukov}},\ }\bibfield  {title} {\bibinfo {title} {{Quantum Phase
  Transition in the Yukawa-SYK Model}},\ }\href
  {https://doi.org/10.1103/PhysRevResearch.2.033084} {\bibfield  {journal}
  {\bibinfo  {journal} {Phys. Rev. Res.}\ }\textbf {\bibinfo {volume} {2}},\
  \bibinfo {pages} {033084} (\bibinfo {year} {2020})},\ \Eprint
  {https://arxiv.org/abs/2005.07205} {arXiv:2005.07205 [cond-mat.str-el]}
  \BibitemShut {NoStop}%
\bibitem [{\citenamefont {Pan}\ \emph {et~al.}(2021)\citenamefont {Pan},
  \citenamefont {Wang}, \citenamefont {Davis}, \citenamefont {Wang},\ and\
  \citenamefont {Meng}}]{PhysRevResearch.3.013250}%
  \BibitemOpen
  \bibfield  {author} {\bibinfo {author} {\bibfnamefont {G.}~\bibnamefont
  {Pan}}, \bibinfo {author} {\bibfnamefont {W.}~\bibnamefont {Wang}}, \bibinfo
  {author} {\bibfnamefont {A.}~\bibnamefont {Davis}}, \bibinfo {author}
  {\bibfnamefont {Y.}~\bibnamefont {Wang}},\ and\ \bibinfo {author}
  {\bibfnamefont {Z.~Y.}\ \bibnamefont {Meng}},\ }\bibfield  {title} {\bibinfo
  {title} {Yukawa-syk model and self-tuned quantum criticality},\ }\href
  {https://doi.org/10.1103/PhysRevResearch.3.013250} {\bibfield  {journal}
  {\bibinfo  {journal} {Phys. Rev. Research}\ }\textbf {\bibinfo {volume}
  {3}},\ \bibinfo {pages} {013250} (\bibinfo {year} {2021})}\BibitemShut
  {NoStop}%
\bibitem [{\citenamefont {Wang}\ \emph {et~al.}(2021)\citenamefont {Wang},
  \citenamefont {Davis}, \citenamefont {Pan}, \citenamefont {Wang},\ and\
  \citenamefont {Meng}}]{PhysRevB.103.195108}%
  \BibitemOpen
  \bibfield  {author} {\bibinfo {author} {\bibfnamefont {W.}~\bibnamefont
  {Wang}}, \bibinfo {author} {\bibfnamefont {A.}~\bibnamefont {Davis}},
  \bibinfo {author} {\bibfnamefont {G.}~\bibnamefont {Pan}}, \bibinfo {author}
  {\bibfnamefont {Y.}~\bibnamefont {Wang}},\ and\ \bibinfo {author}
  {\bibfnamefont {Z.~Y.}\ \bibnamefont {Meng}},\ }\bibfield  {title} {\bibinfo
  {title} {Phase diagram of the spin-$\frac{1}{2}$ yukawa--sachdev-ye-kitaev
  model: Non-fermi liquid, insulator, and superconductor},\ }\href
  {https://doi.org/10.1103/PhysRevB.103.195108} {\bibfield  {journal} {\bibinfo
   {journal} {Phys. Rev. B}\ }\textbf {\bibinfo {volume} {103}},\ \bibinfo
  {pages} {195108} (\bibinfo {year} {2021})}\BibitemShut {NoStop}%
\bibitem [{\citenamefont {Azeyanagi}\ \emph {et~al.}(2018)\citenamefont
  {Azeyanagi}, \citenamefont {Ferrari},\ and\ \citenamefont
  {Massolo}}]{PhysRevLett.120.061602}%
  \BibitemOpen
  \bibfield  {author} {\bibinfo {author} {\bibfnamefont {T.}~\bibnamefont
  {Azeyanagi}}, \bibinfo {author} {\bibfnamefont {F.}~\bibnamefont {Ferrari}},\
  and\ \bibinfo {author} {\bibfnamefont {F.~I.~S.}\ \bibnamefont {Massolo}},\
  }\bibfield  {title} {\bibinfo {title} {Phase diagram of planar matrix quantum
  mechanics, tensor, and sachdev-ye-kitaev models},\ }\href
  {https://doi.org/10.1103/PhysRevLett.120.061602} {\bibfield  {journal}
  {\bibinfo  {journal} {Phys. Rev. Lett.}\ }\textbf {\bibinfo {volume} {120}},\
  \bibinfo {pages} {061602} (\bibinfo {year} {2018})}\BibitemShut {NoStop}%
\bibitem [{\citenamefont {Smit}\ \emph {et~al.}(2021)\citenamefont {Smit},
  \citenamefont {Valentinis}, \citenamefont {Schmalian},\ and\ \citenamefont
  {Kopietz}}]{PhysRevResearch.3.033089}%
  \BibitemOpen
  \bibfield  {author} {\bibinfo {author} {\bibfnamefont {R.~L.}\ \bibnamefont
  {Smit}}, \bibinfo {author} {\bibfnamefont {D.}~\bibnamefont {Valentinis}},
  \bibinfo {author} {\bibfnamefont {J.}~\bibnamefont {Schmalian}},\ and\
  \bibinfo {author} {\bibfnamefont {P.}~\bibnamefont {Kopietz}},\ }\bibfield
  {title} {\bibinfo {title} {Quantum discontinuity fixed point and
  renormalization group flow of the sachdev-ye-kitaev model},\ }\href
  {https://doi.org/10.1103/PhysRevResearch.3.033089} {\bibfield  {journal}
  {\bibinfo  {journal} {Phys. Rev. Research}\ }\textbf {\bibinfo {volume}
  {3}},\ \bibinfo {pages} {033089} (\bibinfo {year} {2021})}\BibitemShut
  {NoStop}%
\bibitem [{\citenamefont {Kim}\ \emph {et~al.}(2020)\citenamefont {Kim},
  \citenamefont {Cao},\ and\ \citenamefont {Altman}}]{PhysRevB.101.125112}%
  \BibitemOpen
  \bibfield  {author} {\bibinfo {author} {\bibfnamefont {J.}~\bibnamefont
  {Kim}}, \bibinfo {author} {\bibfnamefont {X.}~\bibnamefont {Cao}},\ and\
  \bibinfo {author} {\bibfnamefont {E.}~\bibnamefont {Altman}},\ }\bibfield
  {title} {\bibinfo {title} {Low-rank sachdev-ye-kitaev models},\ }\href
  {https://doi.org/10.1103/PhysRevB.101.125112} {\bibfield  {journal} {\bibinfo
   {journal} {Phys. Rev. B}\ }\textbf {\bibinfo {volume} {101}},\ \bibinfo
  {pages} {125112} (\bibinfo {year} {2020})}\BibitemShut {NoStop}%
\bibitem [{\citenamefont {Stanford}(2016)}]{Stanford:2015owe}%
  \BibitemOpen
  \bibfield  {author} {\bibinfo {author} {\bibfnamefont {D.}~\bibnamefont
  {Stanford}},\ }\bibfield  {title} {\bibinfo {title} {{Many-body chaos at weak
  coupling}},\ }\href {https://doi.org/10.1007/JHEP10(2016)009} {\bibfield
  {journal} {\bibinfo  {journal} {JHEP}\ }\textbf {\bibinfo {volume} {10}},\
  \bibinfo {pages} {009}},\ \Eprint {https://arxiv.org/abs/1512.07687}
  {arXiv:1512.07687 [hep-th]} \BibitemShut {NoStop}%
\bibitem [{\citenamefont {Gu}\ \emph {et~al.}(2017)\citenamefont {Gu},
  \citenamefont {Qi},\ and\ \citenamefont {Stanford}}]{Gu:2016oyy}%
  \BibitemOpen
  \bibfield  {author} {\bibinfo {author} {\bibfnamefont {Y.}~\bibnamefont
  {Gu}}, \bibinfo {author} {\bibfnamefont {X.-L.}\ \bibnamefont {Qi}},\ and\
  \bibinfo {author} {\bibfnamefont {D.}~\bibnamefont {Stanford}},\ }\bibfield
  {title} {\bibinfo {title} {{Local criticality, diffusion and chaos in
  generalized Sachdev-Ye-Kitaev models}},\ }\href
  {https://doi.org/10.1007/JHEP05(2017)125} {\bibfield  {journal} {\bibinfo
  {journal} {JHEP}\ }\textbf {\bibinfo {volume} {05}},\ \bibinfo {pages}
  {125}},\ \Eprint {https://arxiv.org/abs/1609.07832} {arXiv:1609.07832
  [hep-th]} \BibitemShut {NoStop}%
\bibitem [{\citenamefont {{Larkin}}\ and\ \citenamefont
  {{Ovchinnikov}}(1969)}]{1969JETP...28.1200L}%
  \BibitemOpen
  \bibfield  {author} {\bibinfo {author} {\bibfnamefont {A.~I.}\ \bibnamefont
  {{Larkin}}}\ and\ \bibinfo {author} {\bibfnamefont {Y.~N.}\ \bibnamefont
  {{Ovchinnikov}}},\ }\bibfield  {title} {\bibinfo {title} {{Quasiclassical
  Method in the Theory of Superconductivity}},\ }\href@noop {} {\bibfield
  {journal} {\bibinfo  {journal} {Soviet Journal of Experimental and
  Theoretical Physics}\ }\textbf {\bibinfo {volume} {28}},\ \bibinfo {pages}
  {1200} (\bibinfo {year} {1969})}\BibitemShut {NoStop}%
\bibitem [{\citenamefont {Maldacena}\ \emph {et~al.}(2016)\citenamefont
  {Maldacena}, \citenamefont {Shenker},\ and\ \citenamefont
  {Stanford}}]{Maldacena:2015waa}%
  \BibitemOpen
  \bibfield  {author} {\bibinfo {author} {\bibfnamefont {J.}~\bibnamefont
  {Maldacena}}, \bibinfo {author} {\bibfnamefont {S.~H.}\ \bibnamefont
  {Shenker}},\ and\ \bibinfo {author} {\bibfnamefont {D.}~\bibnamefont
  {Stanford}},\ }\bibfield  {title} {\bibinfo {title} {{A bound on chaos}},\
  }\href {https://doi.org/10.1007/JHEP08(2016)106} {\bibfield  {journal}
  {\bibinfo  {journal} {JHEP}\ }\textbf {\bibinfo {volume} {08}},\ \bibinfo
  {pages} {106}},\ \Eprint {https://arxiv.org/abs/1503.01409} {arXiv:1503.01409
  [hep-th]} \BibitemShut {NoStop}%
\bibitem [{\citenamefont {Anninos}\ \emph {et~al.}(2016)\citenamefont
  {Anninos}, \citenamefont {Anous},\ and\ \citenamefont {Denef}}]{quiver-1}%
  \BibitemOpen
  \bibfield  {author} {\bibinfo {author} {\bibfnamefont {D.}~\bibnamefont
  {Anninos}}, \bibinfo {author} {\bibfnamefont {T.}~\bibnamefont {Anous}},\
  and\ \bibinfo {author} {\bibfnamefont {F.}~\bibnamefont {Denef}},\ }\bibfield
   {title} {\bibinfo {title} {Disordered quivers and cold horizons},\ }\href
  {https://doi.org/10.1007/JHEP12(2016)071} {\bibfield  {journal} {\bibinfo
  {journal} {Journal of High Energy Physics}\ }\textbf {\bibinfo {volume}
  {2016}},\ \bibinfo {pages} {71} (\bibinfo {year} {2016})}\BibitemShut
  {NoStop}%
\bibitem [{\citenamefont {Anninos}\ \emph {et~al.}(2015)\citenamefont
  {Anninos}, \citenamefont {Anous}, \citenamefont {de~Lange},\ and\
  \citenamefont {Konstantinidis}}]{quiver-2}%
  \BibitemOpen
  \bibfield  {author} {\bibinfo {author} {\bibfnamefont {D.}~\bibnamefont
  {Anninos}}, \bibinfo {author} {\bibfnamefont {T.}~\bibnamefont {Anous}},
  \bibinfo {author} {\bibfnamefont {P.}~\bibnamefont {de~Lange}},\ and\
  \bibinfo {author} {\bibfnamefont {G.}~\bibnamefont {Konstantinidis}},\
  }\bibfield  {title} {\bibinfo {title} {Conformal quivers and melting
  molecules},\ }\href {https://doi.org/10.1007/JHEP03(2015)066} {\bibfield
  {journal} {\bibinfo  {journal} {Journal of High Energy Physics}\ }\textbf
  {\bibinfo {volume} {2015}},\ \bibinfo {pages} {66} (\bibinfo {year}
  {2015})}\BibitemShut {NoStop}%
\bibitem [{\citenamefont {Sachdev}(2015)}]{Sachdev:2015efa}%
  \BibitemOpen
  \bibfield  {author} {\bibinfo {author} {\bibfnamefont {S.}~\bibnamefont
  {Sachdev}},\ }\bibfield  {title} {\bibinfo {title} {{Bekenstein-Hawking
  Entropy and Strange Metals}},\ }\href
  {https://doi.org/10.1103/PhysRevX.5.041025} {\bibfield  {journal} {\bibinfo
  {journal} {Phys. Rev. X}\ }\textbf {\bibinfo {volume} {5}},\ \bibinfo {pages}
  {041025} (\bibinfo {year} {2015})},\ \Eprint
  {https://arxiv.org/abs/1506.05111} {arXiv:1506.05111 [hep-th]} \BibitemShut
  {NoStop}%
\bibitem [{\citenamefont {Murugan}\ \emph {et~al.}(2017)\citenamefont
  {Murugan}, \citenamefont {Stanford},\ and\ \citenamefont
  {Witten}}]{Murugan2017}%
  \BibitemOpen
  \bibfield  {author} {\bibinfo {author} {\bibfnamefont {J.}~\bibnamefont
  {Murugan}}, \bibinfo {author} {\bibfnamefont {D.}~\bibnamefont {Stanford}},\
  and\ \bibinfo {author} {\bibfnamefont {E.}~\bibnamefont {Witten}},\
  }\bibfield  {title} {\bibinfo {title} {More on supersymmetric and 2d analogs
  of the syk model},\ }\href {https://doi.org/10.1007/JHEP08(2017)146}
  {\bibfield  {journal} {\bibinfo  {journal} {Journal of High Energy Physics}\
  }\textbf {\bibinfo {volume} {2017}},\ \bibinfo {pages} {146} (\bibinfo {year}
  {2017})}\BibitemShut {NoStop}%
\bibitem [{\citenamefont {Marcus}\ and\ \citenamefont
  {Vandoren}(2018)}]{Marcus2018ANC}%
  \BibitemOpen
  \bibfield  {author} {\bibinfo {author} {\bibfnamefont {E.}~\bibnamefont
  {Marcus}}\ and\ \bibinfo {author} {\bibfnamefont {S.}~\bibnamefont
  {Vandoren}},\ }\bibfield  {title} {\bibinfo {title} {A new class of syk-like
  models with maximal chaos},\ }\href@noop {} {\bibfield  {journal} {\bibinfo
  {journal} {Journal of High Energy Physics}\ }\textbf {\bibinfo {volume}
  {2019}},\ \bibinfo {pages} {1} (\bibinfo {year} {2018})}\BibitemShut
  {NoStop}%
\bibitem [{\citenamefont {{Sorokhaibam}}(2020)}]{2020JHEP...07..055S}%
  \BibitemOpen
  \bibfield  {author} {\bibinfo {author} {\bibfnamefont {N.}~\bibnamefont
  {{Sorokhaibam}}},\ }\bibfield  {title} {\bibinfo {title} {{Phase transition
  and chaos in charged SYK model}},\ }\href
  {https://doi.org/10.1007/JHEP07(2020)055} {\bibfield  {journal} {\bibinfo
  {journal} {Journal of High Energy Physics}\ }\textbf {\bibinfo {volume}
  {2020}},\ \bibinfo {eid} {55} (\bibinfo {year} {2020})},\ \Eprint
  {https://arxiv.org/abs/1912.04326} {arXiv:1912.04326 [hep-th]} \BibitemShut
  {NoStop}%
\bibitem [{\citenamefont {Murthy}\ \emph {et~al.}(2021)\citenamefont {Murthy},
  \citenamefont {Pandey}, \citenamefont {Esterlis},\ and\ \citenamefont
  {Kivelson}}]{kivelson-2021}%
  \BibitemOpen
  \bibfield  {author} {\bibinfo {author} {\bibfnamefont {C.}~\bibnamefont
  {Murthy}}, \bibinfo {author} {\bibfnamefont {A.}~\bibnamefont {Pandey}},
  \bibinfo {author} {\bibfnamefont {I.}~\bibnamefont {Esterlis}},\ and\
  \bibinfo {author} {\bibfnamefont {S.~A.}\ \bibnamefont {Kivelson}},\ }\href
  {https://doi.org/10.48550/ARXIV.2112.06966} {\bibinfo {title} {A stability
  bound on the $t$-linear resistivity of conventional metals}} (\bibinfo {year}
  {2021})\BibitemShut {NoStop}%
\bibitem [{\citenamefont {Chowdhury}\ and\ \citenamefont
  {Swingle}(2017)}]{PhysRevD.96.065005}%
  \BibitemOpen
  \bibfield  {author} {\bibinfo {author} {\bibfnamefont {D.}~\bibnamefont
  {Chowdhury}}\ and\ \bibinfo {author} {\bibfnamefont {B.}~\bibnamefont
  {Swingle}},\ }\bibfield  {title} {\bibinfo {title} {Onset of many-body chaos
  in the $o(n)$ model},\ }\href {https://doi.org/10.1103/PhysRevD.96.065005}
  {\bibfield  {journal} {\bibinfo  {journal} {Phys. Rev. D}\ }\textbf {\bibinfo
  {volume} {96}},\ \bibinfo {pages} {065005} (\bibinfo {year}
  {2017})}\BibitemShut {NoStop}%
\bibitem [{\citenamefont {Tikhanovskaya}\ \emph {et~al.}(2022)\citenamefont
  {Tikhanovskaya}, \citenamefont {Sachdev},\ and\ \citenamefont
  {Patel}}]{PhysRevLett.129.060601}%
  \BibitemOpen
  \bibfield  {author} {\bibinfo {author} {\bibfnamefont {M.}~\bibnamefont
  {Tikhanovskaya}}, \bibinfo {author} {\bibfnamefont {S.}~\bibnamefont
  {Sachdev}},\ and\ \bibinfo {author} {\bibfnamefont {A.~A.}\ \bibnamefont
  {Patel}},\ }\bibfield  {title} {\bibinfo {title} {Maximal quantum chaos of
  the critical fermi surface},\ }\href
  {https://doi.org/10.1103/PhysRevLett.129.060601} {\bibfield  {journal}
  {\bibinfo  {journal} {Phys. Rev. Lett.}\ }\textbf {\bibinfo {volume} {129}},\
  \bibinfo {pages} {060601} (\bibinfo {year} {2022})}\BibitemShut {NoStop}%
\end{thebibliography}%

\end{document}